\documentclass[12pt]{article}
  
\usepackage[a4paper, total={6.25in, 9.5in}]{geometry}
\usepackage[super,square]{natbib}
\usepackage{doc}
\usepackage{subcaption}
\usepackage{url}
\usepackage{sectsty}
\sectionfont{\fontsize{16}{15}\selectfont}
\subsectionfont{\fontsize{16}{15}\selectfont}
\allsectionsfont{\mdseries}
\usepackage{graphicx}
\usepackage{epstopdf}
\title{\Large An Investigation into the $HfO_2/Si$ Interface: Materials Science Challenges and their Effects on MOSFET Device Performance}
\author{Aditya Muralidharan}
\date{}

\begin{document}

\maketitle

\abstract{
Since the 1960's when Gordon Moore proposed that the transistor density in our electronic devices should double every two years while the cost is halved, the semiconductor industry has taken this statement to heart. Over the last few decades, no other industry has seen growth even comparably close to that experienced by the semiconductors industry. This has all been made possible by the unbroken string of ingenious breakthroughs by brilliant minds that have been working tirelessly to shrink down transistors. The latest of which is the use of high-k dielectrics and a return to metal gates combined with $3D$-transistor architectures. This has been the enabling technology for the transition from the $90\;nm$ node to the $45\;nm$ node, allowing us to shrink our transistors further without losing additional gate control. The fundamental reason for using a high-k gate dielectric compared to $SiO_2$ is that shrinking our gate oxide further, which is already at a few angstroms, is no longer a feasible option to gain additional gate control. High-k dielectric overcome this by exploiting the fundamental physics of capacitors and the materials science of dielectrics to provide a viable option to increase gate control without the need for successively thinner gate oxides. Hafnium Oxide $(HfO_2)$ is the most studied and popular of such materials. Its high dielectric constant $\sim 16-25$ and interface stability with silicon at operating temperatures make it an ideal candidate for use in current $CMOS$ technology. Intel has been using $HfO_2$ for their logic technology foundries for its $32\;nm$ node and Micron its partner in memory has been using $HfO_2$ for its $DRAM$ access transistors and capacitors for the $50\;nm$ node, all since the early 2010's. Despite its deceivingly simple appearance, the processes involved in the fabrication of such high-k $HfO_2/Si$ interfaces are full of process subtleties and fabrication nuances. One of the primary reasons for this is the formation of a $SiO_2$ interlayer after deposition at process temperatures $(\sim 600^oC-1000^oC)$ which degrades our final transistors gate control and ultimately the device characteristics.

In this term paper we hope to explore the physics and materials science of these high-k $HfO_2/Si$ interfaces, discussing the challenges and ways to overcome them when it comes to its actual fabrication, and how this ultimately affects our device performance. This paper is broadly divided into three sections where we look at the thermodynamics, reaction kinetics and atomic diffusion kinetics, after which we look at the effects of these parameters on device performance and fabrication.
}

\pagebreak

\pagebreak

\pagebreak

\pagebreak

%
%
\section*{\normalfont Introduction }
\label{introduction}

\begin{figure}[b]
	\centering
	\begin{subfigure}[b]{0.45\textwidth}
		\centering
		\includegraphics[height=0.9in]{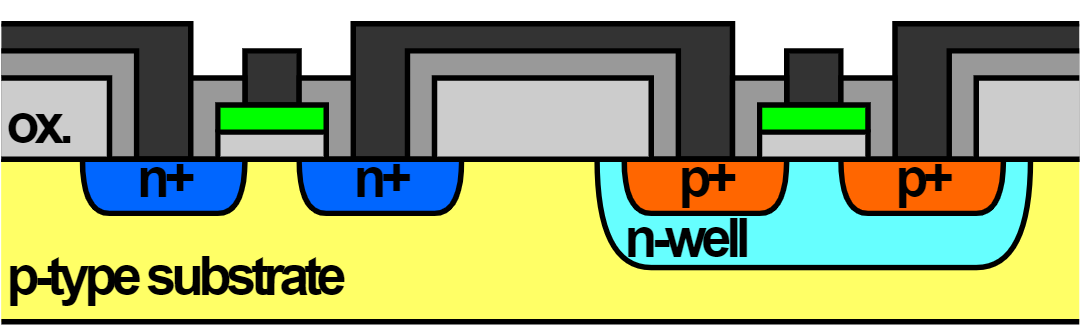}
		\caption{{}}
	\end{subfigure}
	\begin{subfigure}[b]{0.45\textwidth}
	\centering
	\includegraphics[height=1.5in]{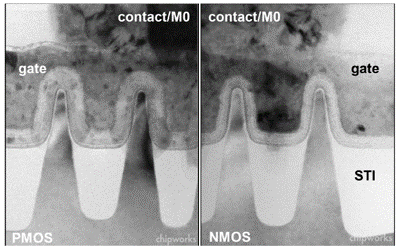}
	\caption{{}}
	\end{subfigure}

	\caption{\textbf{\textit{(a)}} Schematic of a CMOS transistor pair. 
			 \textbf{\textit{(b)}} Image showing Intel's $45\;nm$ node CMOS using high-k dielectrics and metal-gates on strained silicon $FinFET$'s.\cite{james}
	}
\end{figure} 

\begin{figure}[t]
	\centering
	\begin{subfigure}[b]{0.45\textwidth}
		\centering
		\includegraphics[height=2in]{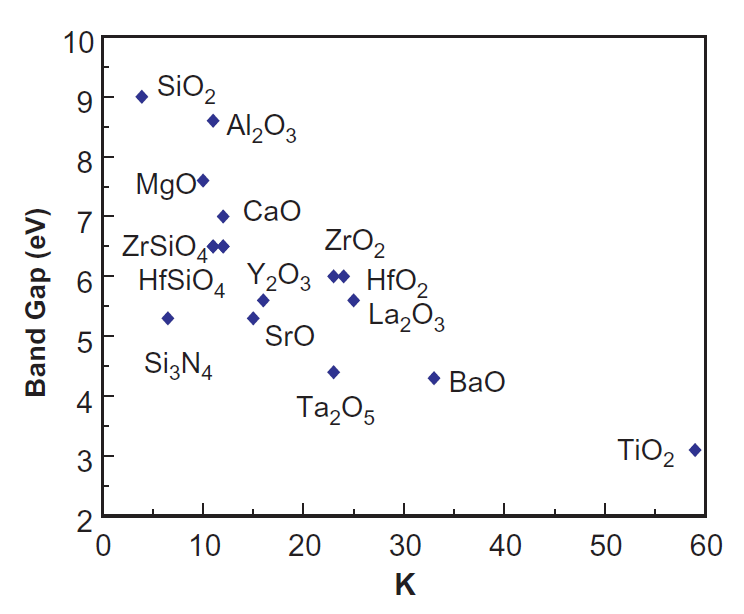}
		\caption{{}}
	\end{subfigure}
	\begin{subfigure}[b]{0.45\textwidth}
		\centering
		\includegraphics[height=2in]{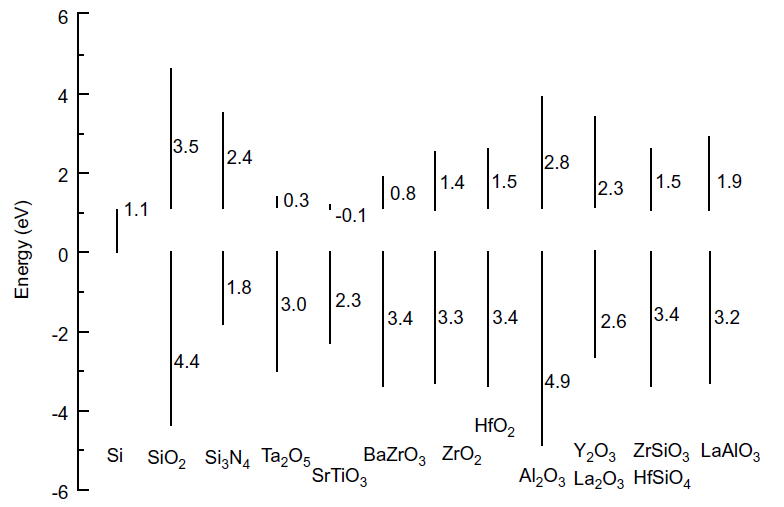}
		\caption{{}}
	\end{subfigure}
	
	\caption{\textbf{\textit{(a)}} Potential high-k dielectrics plotted by their band-gap and dielectric constants. 
		\textbf{\textit{(b)}} Band-offsets of various high-k oxides with silicon. 
	}
\end{figure} 

Moore's law has been the guiding force for the semiconductors industry since the last few decades, enabling it to grow at unprecedented rates unmatched by any other industry.\cite{james}\cite{chang} It is no understatement to say that this urge to double the transistor density while roughly halving the cost every two years has driven the enormous progress seen in the current information age\cite{james}. Complementary MOSFET technology or CMOS is at the heart of all this development, offering many distinct advantages over its competitors.\cite{james}\cite{engstrom}\cite{gusev} Especially zero static power consumption,  which makes large circuits feasible and scalable.\cite{gusev} The ability to achieve high logic density on $IC$'s while maintaining ease in integration also played a major role in CMOS triumphing.\cite{james}\cite{gusev} A unique feature of CMOS, is the use of both NMOS and PMOS devices as complementary elements of a singular logic circuit thereby lending it it's name.\cite{james}\cite{gusev} This necessitates that the fabrication of such devices take place on a shared substrate with diffusion well regions for a PMOS device always next to a NMOS device or vice-versa, as seen in Figure 1.\cite{james}\cite{chang} This adds to the complexity in fabrication but with modern techniques is easily achievable and makes the resulting circuits much more compact compared to other logic paradigms.\cite{james}\cite{chang} However, in recent years, as we approach atomic scale dimensions the continued miniaturization of transistor has become increasingly troublesome. Quantum effects such as $DIBL$, channel-length modulation and the other so called short channel effects dominate as one encroaches into further decreasing length scales.\cite{james}\cite{engstrom} Furthermore, fabrication limitations on creating atomically thin gate oxides reliably, further drives up the cost.\cite{chang}\cite{gougam}\cite{kim} To combat this, engineers have looked to other routes to miniaturization. \cite{chang}\cite{kim}

Intel in 2007 announced a breakthrough, their use of high-k dielectrics and metal gates combined with FinFET technology could provide a never before seen combination of performance and reliability.\cite{james} These materials have a higher dielectric constant, allowing us to fabricate thicker physical films while achieving an oxide electrical thickness $(EOT)$ equivalent to  thinner $SiO_2$ oxides by suppressing quantum mechanical effects.\cite{james}\cite{chang}\cite{kim} Initially, only the dielectric constant of oxides was considered a deciding criteria.\cite{chang}\cite{robertson} This was however an incorrect approach, as the polarizability of the film depended not only on the materials used; most commonly transition metal oxides like $ZrO_2$, $TiO_2$ and $HfO_2$; but also on the films material properties like its crystallinity and crystallographic phase, electrical properties such as band-gap, chemical stability at the interface and overall thermodynamic stability.\cite{chang} This was the enabling technology that pushed Moore's law to the $45\;nm$ node and beyond. From then on it was a race to see who could commercialize this technology first.\cite{james} Major players like Samsung, TSMC, UMC and Globalfoundries joined the race.\cite{james} This however, was not without its own challenges, many high-k dielectrics were considered as potential candidates for commercial device fabrication but a higher polarizability necessarily leads to weaker bonding thereby degrading the electrical stability and breakdown fields achievable.\cite{chang} This makes many high-k materials with extremely high dielectric constants like $SrTiO_3$ and $Ta_2O_3$ unsuitable for use in CMOS technology.\cite{chang} Consequently, weak bonding also makes the interface of these materials with silicon highly unstable causing undesired reactions with silicon at the interface.\cite{chang}

It is therefore helpful to build a set of criteria that the high-k dielectric must satisfy to make it suitable for CMOS applications.\cite{chang} Promising high-k materials should ideally have a dielectric constant between $10-30$ and a band-gap of $5\;eV$ and maintain a band offset of at least $1\;eV$ with the silicon substrate.\cite{chang} Additionally, to facilitate process feasibility the dielectric should be thermally stable at process temperature exceeding $1000\;K$ for more than $90\sec$.\cite{chang}
This leaves the primary candidate for the high-k dielectric to be used as $HfO_2$, having a dielectric constant of $\sim25$, depending on the crystallographic phase, and a relatively large bandgap of $5.7\;eV$.\cite{chang} The larger heat of formation of $HfO_2$ compared to $SiO_2$ makes it stable on the substrate at room temperature and device operating temperatures, however at process temperatures forms a $SiO_2$ or $HfSiO_4$ interlayer which degraded device performance.\cite{chang}  At an operating voltage of $1-1.5\;V$ the leakage current through a $HfO_2$ gate oxide was several orders of magnitude lower than $SiO_2$ at the same $EOT$.\cite{chang}\cite{robertson}

In this term paper we look at the various materials science challenges associated with the fabrication of an electrically and thermodynamically stable $HfO_2/Si$ interface and explore the mechanism of interlayer formation.\cite{chang}\cite{gutowski} We dive into the  thermodynamics and kinetics at the interface and take a critical look at the factors that enable $SiO_2$ formation and how it can be mitigated.\cite{chang}\cite{gutowski}\cite{shin}\cite{shin2} Finally, we look at the device performance of the gate stack compare it to a device with $SiO_2$ interlayer and analyse how it degrades device performance.\cite{gusev}\cite{miyata}\cite{gusev2}

\section*{\normalfont The $HfO_2-$Silicon Interface }

\begin{figure}[t]
	\centering
	\begin{subfigure}[b]{1\textwidth}
		\centering
		\includegraphics[height=2in]{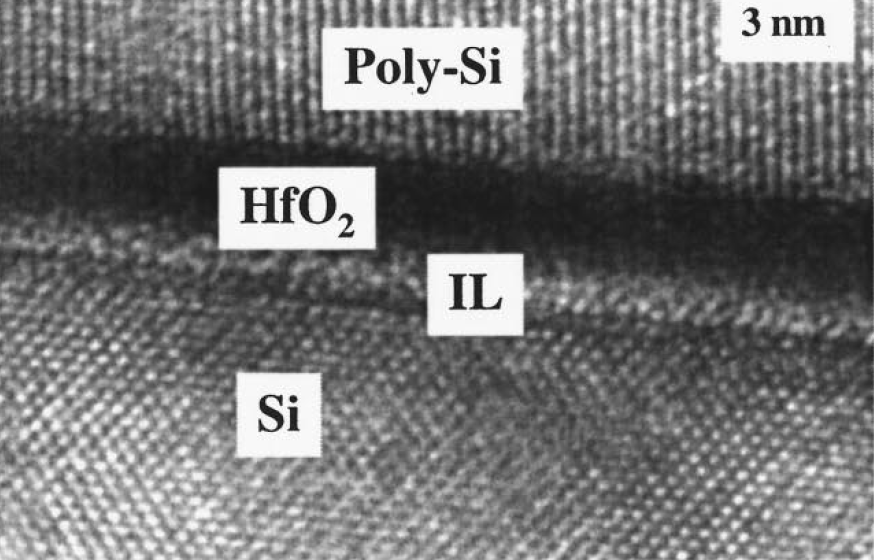}
		\caption{{}}
	\end{subfigure}
	
	\caption{\textbf{\textit{(a)}} $HRTEM$ of a $HfO_2$-Silicon interface annealed at $1025^oC$ for $20$ seconds, showing the $SiO_2$ interlayer. \cite{gutowski}
	}
\end{figure} 

The oxide-silicon interface on a $MOSFET$ transistor is a region of critical importance to the electrical performance of the device.\cite{chang}\cite{caymax}\cite{chiu} The primary reason why silicon is used to fabricate almost all integrated circuits despite being an inferior semiconductor compared to alternatives like germanium, is the ease of growing a chemically and electrically stable oxide on its surface.\cite{chang} Using high-k dielectrics as gate oxides is not a simple proposition, and introduces challenges in fabrication and device performance.\cite{chang}\cite{banyay} The $HfO_2-$silicon interface has been extensively studied and the presence of a $SiO_2$ interlayer is well established.\cite{chang}\cite{gutowski}\cite{banyay} The interlayer is primarily amorphous and the trap charge density is dependent on the growth conditions of the $HfO_2$ film.\cite{chang}\cite{robertson2}\cite{zhang}\cite{singh} Figure 3 shows the interlayer in a $HfO_2$ film annealed at a temperature of $1025^oC$ for $20$ seconds.\cite{gutowski} The interlayer is known to reduce the effective dielectric constant of the film, effectively increasing the $EOT$ of the film.\cite{chang}\cite{robertson} \cite{chiu}\cite{copel}This reduces the gate control on our channel, thus, degrading the performance of our device.\cite{chang}\cite{robertson}

In this section we first take a closer look at the properties of $HfO_2$ and its advantages compared to other high-k dielectrics, after which we take a deeper dive into the thermodynamics of the interface and how it affects the chemical and crystallographic properties of the oxide. We then try to justify the formation of the interlayer and look at the reaction kinetics at play. Finally we try to give an atomistic picture of the underlying mechanism and ways to mitigate interlayer formation.

\subsection*{\normalfont $HfO_2$}

\subsubsection*{\normalfont Structural Properties}

\begin{figure}[t]
	\centering
	\begin{subfigure}[b]{0.3\textwidth}
		\centering
		\includegraphics[height=1.5in]{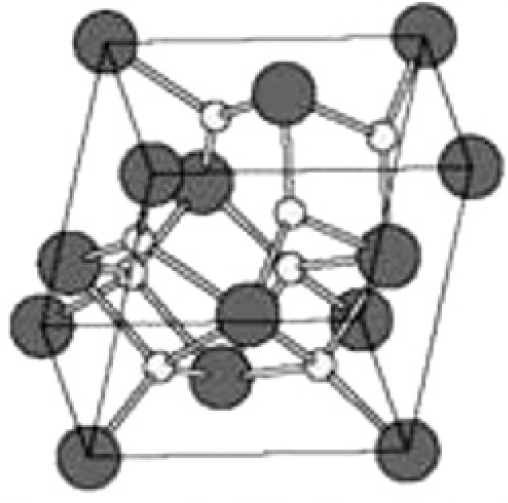}
		\caption{{}}
	\end{subfigure}
	\begin{subfigure}[b]{0.3\textwidth}
	\centering
	\includegraphics[height=1.5in]{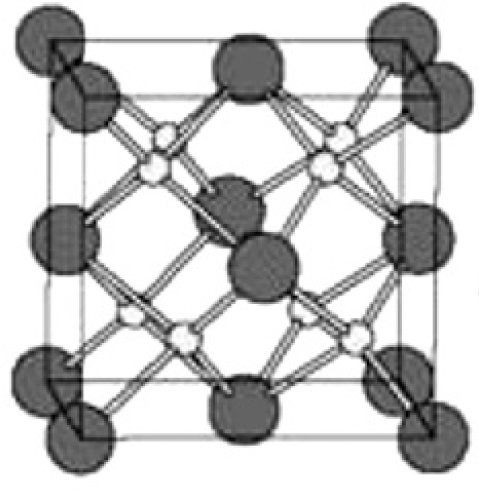}
	\caption{{}}
\end{subfigure}
	\begin{subfigure}[b]{0.3\textwidth}
	\centering
	\includegraphics[height=1.5in]{fig4_2.png}
	\caption{{}}
\end{subfigure}
	
	\caption{\textbf{\textit{(a)}} Monoclinic $HfO_2\;(a=5.1$\AA,$b=5.17$\AA,$c=5.26$\AA$)$\cite{chang}
		\textbf{\textit{(b)}} Tetragonal $HfO_2\;(a=5.03$\AA$,b=5.04$\AA$,c=5.18$\AA$)$\cite{chang}
		\textbf{\textit{(c)}} Cubic $HfO_2\;(a=5.06$\AA$,b=5.06$\AA$,c=5.06$\AA$)$\cite{chang}
	}
\end{figure} 

Group $IV-B$ elements, namely $Hf$, $Zr$, $Ti$ and their oxides have gained a lot of attention in recent years for their use in $CMOS$ technology as high-k dielectrics.\cite{chang}\cite{robertson} Among these Hafnium Oxide $(HfO_2)$ shows the most promise due to its high dielectric constant $(k\sim22-25)$, large band-gap $(\sim5.7\;eV)$, high breakdown field $(3.9-6.7\;MV\;cm^{-1})$, and thermodynamics stability $\Delta H_{for}=-271Kcal \;mol^{-1}$.\cite{chang}\cite{kim3} The crystallographic phase of the film plays an important role in determining the properties of the dielectric film.\cite{chang}\cite{kim3} $HfO_2$ forms a monoclinic phase at room temperature, with the $Hf$ atom exhibiting a $7$ fold co-ordination being surrounded by $O$ atoms, as seen in Figure 4.\cite{chang} The monoclinic structure has the lowest free energy of formation and the largest volume at room temperature.\cite{chang} Upon annealing at $1024^oC$ it undergoes a phase transformation to its tetragonal phase, and a cubic phase which is only formed when annealed at $2422^oC$.\cite{chang}

As deposited thin films of $HfO_2$, however, are mostly polycrystalline often exhibiting multiple crystallographic phases.\cite{chang}\cite{zhang} Cubic and tetragonal phases of $HfO_2$ have been observed along with the monoclinic phase at room temperature, primarily at the interface due to strains.\cite{chang}\cite{zhang} The tetragonal phase also forms a lattice matched interface $($mismatch$<5\%)$ with $(110)$ oriented silicon minimizing the dangling bonds at the interface.\cite{chang}\cite{kolkovsky} Films obtained by $ALD$ at $500^oC$ also exhibit an orthorhombic phase, whereas when performed at $325^oC$ weak peaks of the tetragonal phase were also observed.\cite{chang}\cite{gougam}\cite{zhang} Cubic nano-crystallites of $HfO_2$ were also observed at growth performed at $900^oC$.\cite{chang}\cite{xu}\cite{senzaki} These characteristics are closely determined by the processing conditions under which the films are grown, complicating the formation of our gate oxide.\cite{chang}\cite{senzaki} Thus, it is preferable to grow perfectly amorphous or single crystalline $HfO_2$ films for $CMOS$ applications.\cite{chang}\cite{kolkovsky}\cite{sayan} As will be discussed in following sections this is also done to ensure that oxygen diffusion at the interface can be minimized to mitigate the growth of the interlayer.\cite{chang}\cite{copel}\cite{zafar}

\subsubsection*{\normalfont Electrical Properties}

\begin{figure}[t]
	\centering
	\begin{subfigure}[b]{1\textwidth}
		\centering
		\includegraphics[height=3in]{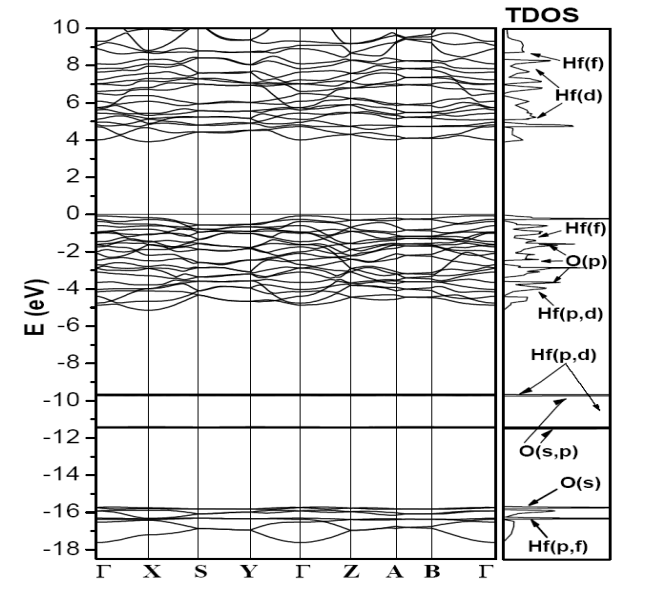}
		\caption{{}}
	\end{subfigure}
	
	\caption{\textbf{\textit{(a)}} Band-structure of $m-HfO_2$.\cite{garcia}
	}
\end{figure} 

\begin{figure}[b]
	\centering
	\begin{subfigure}[b]{0.22\textwidth}
		\centering
		\includegraphics[height=1.4in]{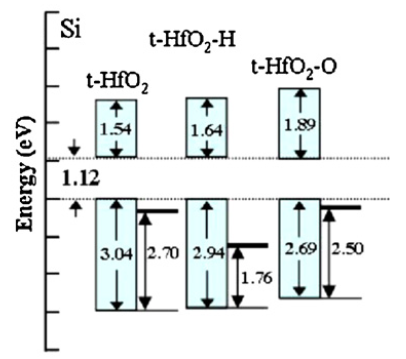}
		\caption{{}}
	\end{subfigure}
	\begin{subfigure}[b]{0.22\textwidth}
		\centering
		\includegraphics[height=1.4in]{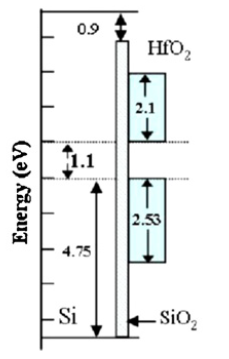}
		\caption{{}}
	\end{subfigure}
	\begin{subfigure}[b]{0.22\textwidth}
		\centering
		\includegraphics[height=1.4in]{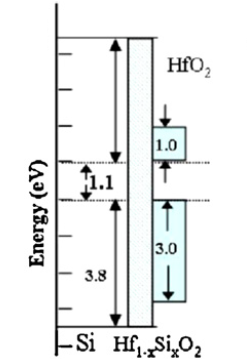}
		\caption{{}}
	\end{subfigure}
	\begin{subfigure}[b]{0.22\textwidth}
		\centering
		\includegraphics[height=1.4in]{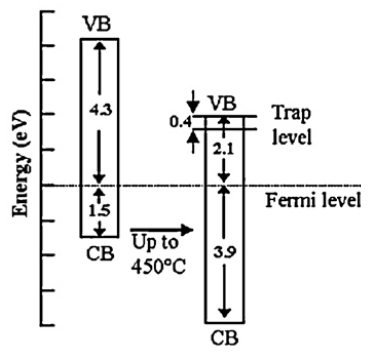}
		\caption{{}}
	\end{subfigure}
	
	\caption{\textbf{\textit{(a)}} Band-offset of $HfO_2$ with $H$ or $O$ interface passivation compared to no interface passivation.\cite{chang}
		\textbf{\textit{(b)}} Band alignments in $HfO_2/SiO_2/Si$ heterostructure.\cite{chang}
		\textbf{\textit{(c)}} Band alignments in $HfO_2/Hf_{1-x}Si_xO_2/Si$ heterostructure.\cite{chang}
		\textbf{\textit{(d)}} Change in the band alignment due to annealing at different temperature.\cite{chang}
	}
\end{figure} 

\begin{figure}[t]
	\centering
	\begin{subfigure}[b]{1\textwidth}
		\centering
		\includegraphics[height=2in]{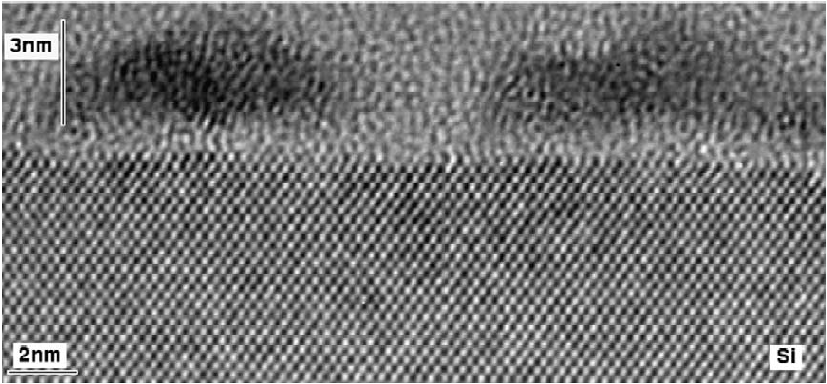}
		\caption{{}}
	\end{subfigure}
	
	\caption{\textbf{\textit{(a)}} $HfO_2$ films directly deposited on a $(110)$ silicon substrate freshly after $HF$ treatment. One can notice the clear lack of an oxide interlayer.\cite{gusev}
	}
\end{figure} 

The electrical properties of $HfO_2$ have been well studied in literature.\cite{chang}\cite{chiu}\cite{kim3}\cite{garcia} Exhibiting a bandgap of $5.7\;eV$, and a band-offset of around $1.5\;eV$ with silicon, $HfO_2$ has been determined to be a suitable material for use as a high-k dielectric on silicon for a variety of reasons.\cite{chang} The primary reason was noted to be the $4f-5d$ hybridization of the hafnium atoms with the $2s$ and $2p$ subshells of oxygen, resulting in a valence band consisting of three separate sub bands separated by ionic gaps, namely the $O(2d)-Hf(4f\;,\;5d)$ state, $O(2s)-Hf(4f)$ state and the $Hf(5p)$ state.\cite{chang}\cite{garcia} This results in the aforementioned bandgap and the existence of a light-hole $(0.3m_o)$ band and a heavy-hole $(8.3m_o)$ band, with the electron effective mass in the range of $0.7m_o-2m_o$.\cite{chang}\cite{garcia} The band diagram of monoclinic $HfO_2$ is shown in Figure 5.\cite{garcia}

The characteristics of oxygen deficient $HfO_2$ is also of great interest to us, primarily due to the fact that it shows a slight increase in the conduction band density of states.\cite{chang}\cite{copel}\cite{almeida} Extended defects in general can affect the electronic behaviour of the dielectric film, causing issues with device reliability.\cite{chang}\cite{robertson}\cite{robertson2}\cite{mckenna}\cite{kang} $DFT$ studies of the oxygen vacancy diffusion in $HfO_2$ have shown that both neutral and positive vacancies preferentially segregate at grain boundaries, providing percolation path for electrons.\cite{chang}\cite{mckenna}\cite{mueller} Positive vacancies have a diffusion activation energy of $\sim0.7\;eV$, diffusing freely in the bulk until their inevitable accumulation at grain boundaries.\cite{zafar} At the interface these positive vacancies interact with silicon atoms gaining electrons from them eventually leading to larger defect cluster and interlayer formation.\cite{chang}\cite{zafar}\cite{mckenna}

Composition variation in the $HfO_2$ also has an important impact on the properties of the film at the interface.\cite{chang}\cite{copel} First principle calculations have revealed the partial occupancy of the $Hf(5d)$ orbitals results in dangling bonds, becoming one of the culprits for the high band offset upon formation of an interlayer with silicon explored in more detail in the following section.\cite{chang}\cite{robertson2} The band offset of the interface is therefore highly dependent on the quality of the interface, with a theoretical range from $2.69\;eV-3.04\;eV$ for the conduction band and $1.54V-1.89V$ for the valence band.\cite{chang} With heterostructures such as $HfO_2/HfSiO_x/Si$ the band offsets are $4.75\;eV$ for $SiO_x$ and $2.53\;eV$ for $HfO_2$.\cite{chang} Annealing of the interface has been known to affect the band alignment at the interface, primarily due to chemical modification and changes in crystallography.\cite{chang}\cite{gougam}\cite{zhang} These effects are summarized in Figure 6.\cite{chang}

The breakdown characteristics of the oxide determine the reliability of the oxide, and ultimately the fabricated device.\cite{chang}\cite{chiu} The mechanism of breakdown is usually by an avalanche process, beginning at a single weak point; usually a defect; resulting in a weak and localized conduction path between the gate and the substrate, called soft breakdown.\cite{chang} This eventually leads to joule heating along the pathway, ultimately leading to hard breakdown.\cite{chang} Oxide breakdown is usually described in terms of a Weibull distribution of the form $F(t)=1-\exp{\frac{-\beta t}{\alpha}}$, with $\beta$ being the Weibull modulus and $\alpha$ being the characteristic time.\cite{chang} With proper engineering and process optimization an ultra-thin $HfO_2$ film, has been shown to meet the performance and reliability criteria for use as gate dielectric.\cite{chang}\cite{caymax} The deposition technique used for fabrication has significant consequences on the crystallographic structure, defect density, interface states and band alignment, further explored in following sections.\cite{chang}\cite{banyay}\cite{zhang}\cite{senzaki}

\subsection*{\normalfont Thermodynamic \& Chemical Stability}

\begin{table}
	\centering
	\caption{Possible reactions occurring at the high-k silicon interface\cite{gutowski}}
	\begin{tabular}{lllccc}
		&Reaction & $\Delta H_{DFT}$ & $\Delta H_{exp}$ &  $\Delta H_{DFT}$& $\Delta G_{exp}$\\
		&        & $(0K)$         & $(298K)$       &  $(1000K)$     & $(1000K)$      \\
	    &  & $\frac{KJ}{mol}$ & $\frac{KJ}{mol}$ &  $\frac{KJ}{mol}$ & $\frac{KJ}{mol}$ \\
		\hline
	H1	&$Si + HfO_2\rightarrow Hf +SiO_2 $ 	   	  & 227.9 &  -   &  -   &- \\
	H2	&$2Si + HfO_2\rightarrow HfSi +SiO_2 $     & 71.4  & 92.1 &  -   &- \\
	H3	&$3Si + HfO_2\rightarrow HfSi_2 +SiO_2 $   & 68.1  &  -   & 	-   &-\\
	H4	&$2Si + HfO_2\rightarrow HfSi +HfSiO_4 $   & 59.0  &  -   &  -   &- \\
	H5	&$3Si + HfO_2\rightarrow HfSi_2 +HfSiO_4 $ & 55.7  &  -	 &  -   &- \\
	H6	&$Si + 2HfO_2\rightarrow Hf +HfSiO_4 $ 	  & 215.6 &  -   &  -   &- \\
	H7	&$SiO_2 + HfO_2\rightarrow HfSiO_4 $       & -12.4 &  -   &  -   &- \\
		\hline
	Z1	&$Si + ZrO_2\rightarrow Zr +SiO_2 $ 	   	  & 165.8  		 & 186.6 & 185.6 & 177.7 \\
	Z2	&$2Si + ZrO_2\rightarrow ZrSi +SiO_2 $     & -8.5 		 & -2.4  & -4.9  & -11.0\\
	Z3	&$3Si + ZrO_2\rightarrow ZrSi_2 +SiO_2 $   & -13.5 		 & 5.7   & 1.9   & 3.3\\
	Z4	&$2Si + ZrO_2\rightarrow ZrSi +ZrSiO_4 $   & -25.3 		 & -27.9 & -22.0 & -17.5\\
	Z5	&$3Si + ZrO_2\rightarrow ZrSi_2 +ZrSiO_4 $ & -30.3 		 & -19.8 & 15.2  & -3.2\\
	Z6	&$Si + 2ZrO_2\rightarrow Zr +ZrSiO_4 $ 	  &  149.0		 & 161.1 & 168.5 & 171.2\\
	Z7	&$SiO_2 + ZrO_2\rightarrow ZrSiO_4 $       & -16.8        & -25.5 & -17.1 & -6.5\\
	\end{tabular}
\end{table}

\begin{figure}[t]
	\centering
	\begin{subfigure}[b]{0.3\textwidth}
		\centering
		\includegraphics[height=1.5in]{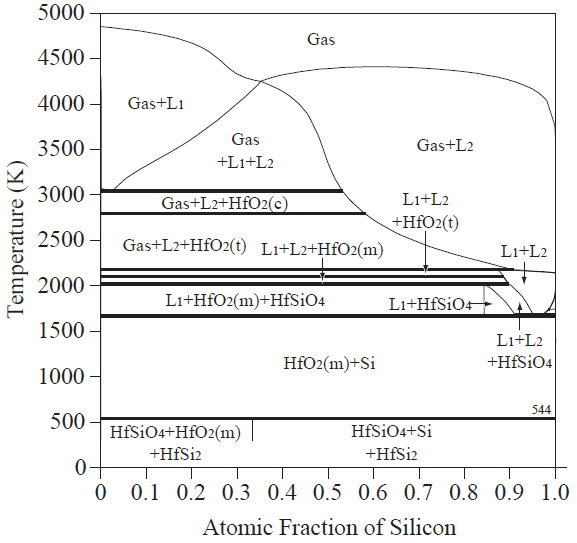}
		\caption{{}}
	\end{subfigure}
	\begin{subfigure}[b]{0.3\textwidth}
		\centering
		\includegraphics[height=1.5in]{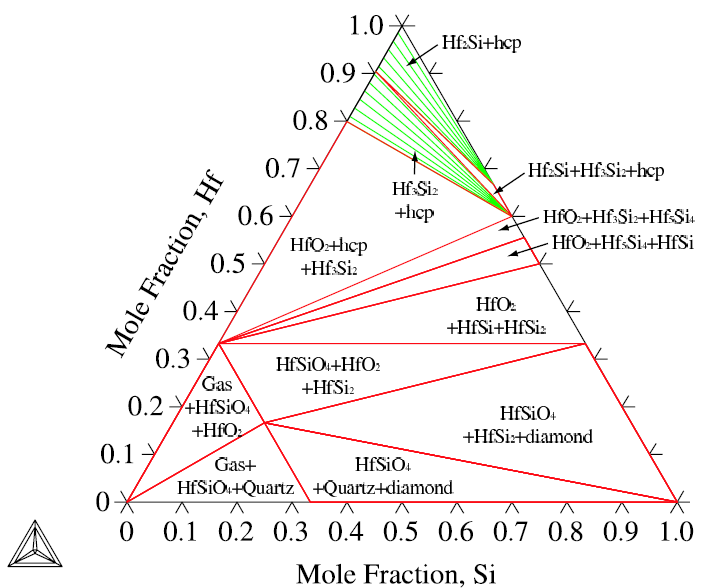}
		\caption{{}}
	\end{subfigure}
	\begin{subfigure}[b]{0.3\textwidth}
		\centering
		\includegraphics[height=1.5in]{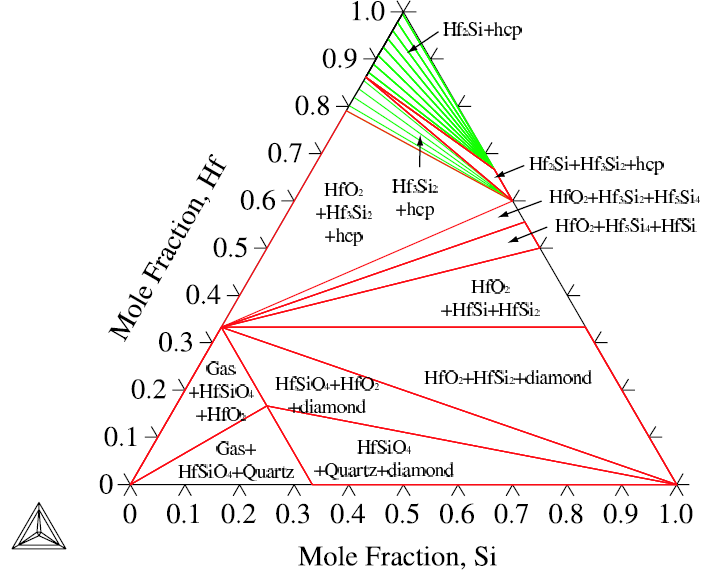}
		\caption{{}}
	\end{subfigure}

	\begin{subfigure}[b]{0.3\textwidth}
		\centering
		\includegraphics[height=1.5in]{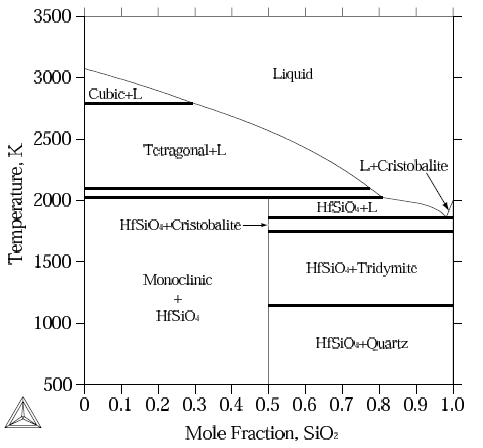}
		\caption{{}}
	\end{subfigure}
	\begin{subfigure}[b]{0.3\textwidth}
		\centering
		\includegraphics[height=1.5in]{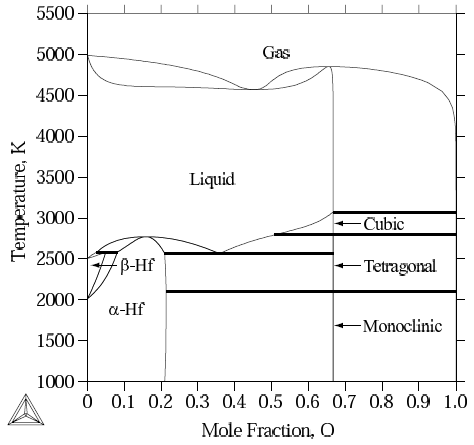}
		\caption{{}}
	\end{subfigure}
	\begin{subfigure}[b]{0.3\textwidth}
		\centering
		\includegraphics[height=1.5in]{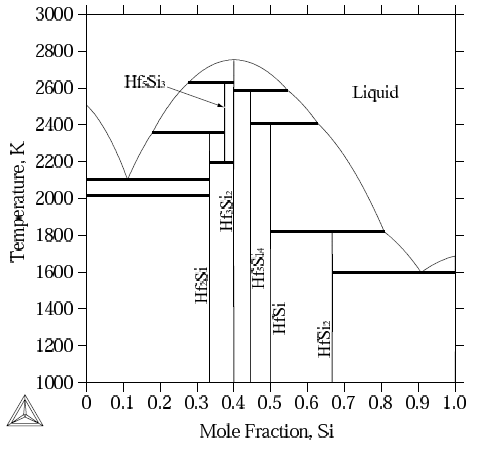}
		\caption{{}}
	\end{subfigure}

	\begin{subfigure}[b]{0.45\textwidth}
		\centering
		\includegraphics[height=1.5in]{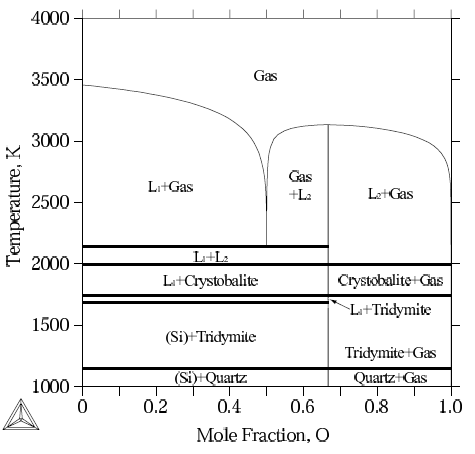}
		\caption{{}}
	\end{subfigure}
	\begin{subfigure}[b]{0.45\textwidth}
		\centering
		\includegraphics[height=1.5in]{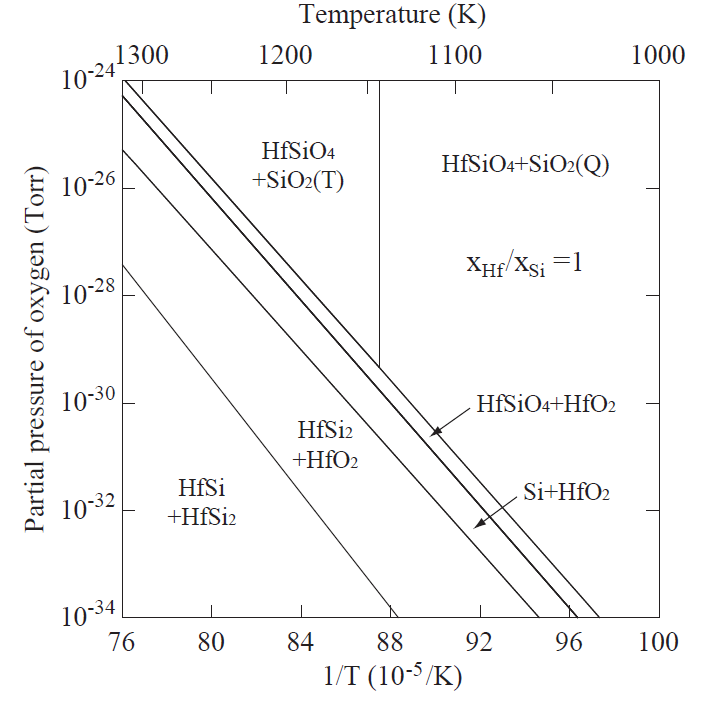}
		\caption{{}}
	\end{subfigure}

	\caption{\textbf{\textit{(a)}} $HfO_2-Si$ Isopleth
		\textbf{\textit{(b)}} Ternary phase diagram of the $Hf-Si-O$ system at $500K$\cite{shin}
		\textbf{\textit{(c)}} Ternary phase diagram of the $Hf-Si-O$ system at $1000K$\cite{shin2}
		\textbf{\textit{(d)}} $HfO_2-SiO_2$ binary phase diagram.\cite{shin2}
		\textbf{\textit{(e)}} $Hf-O$ binary phase diagram.\cite{shin2}
		\textbf{\textit{(f)}} $Hf-Si$ binary phase diagram.\cite{shin2}
		\textbf{\textit{(g)}} $Si-O$ binary phase diagram.\cite{shin2}
		\textbf{\textit{(h)}} $p_o$ vs $T$ phase diagram at the interface.\cite{shin}
	}
\end{figure} 

\begin{figure}[t]
	\centering
	\begin{subfigure}[b]{0.3\textwidth}
		\centering
		\includegraphics[height=1.8in]{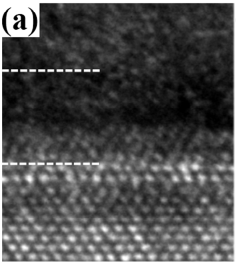}
		\caption{{}}
	\end{subfigure}
	\begin{subfigure}[b]{0.3\textwidth}
	\centering
	\includegraphics[height=1.8in]{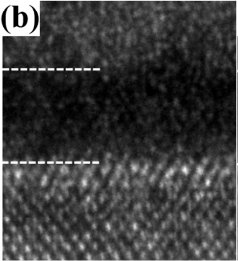}
	\caption{{}}
\end{subfigure}
	\begin{subfigure}[b]{0.3\textwidth}
	\centering
	\includegraphics[height=1.8in]{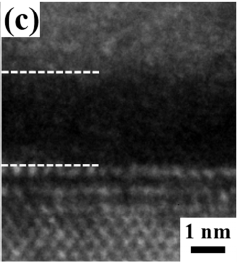}
	\caption{{}}
\end{subfigure}

	\begin{subfigure}[b]{0.45\textwidth}
	\centering
	\includegraphics[height=1.5in]{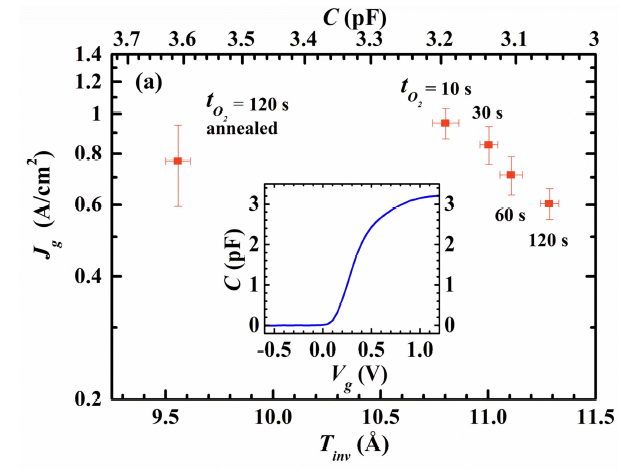}
	\caption{{}}
\end{subfigure}
	\begin{subfigure}[b]{0.45\textwidth}
	\centering
	\includegraphics[height=1.4in]{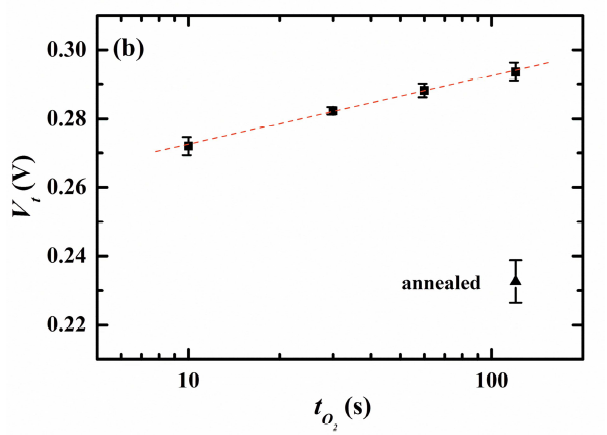}
	\caption{{}}
\end{subfigure}

	\caption{ $HfO_2$ films directly deposited on a $(110)$ substrate using $PVD$ subjected to an oxidation time of \textbf{\textit{(a)}} $120\;s$\cite{jamison} \textbf{\textit{(b)}} $10\;s$\cite{jamison} \textbf{\textit{(c)}} $120\;s$ but with an additional anneal step at $750^oC$ for $5\;$min\cite{jamison}
		\textbf{\textit{(e)}} Leakage characteristics and gate capacitance of the films.\cite{jamison}
		\textbf{\textit{(f)}} Threshold voltage as a function of the oxidation time.\cite{jamison}
	}
\end{figure}

\begin{figure}[b]
	\centering
	\begin{subfigure}[b]{0.14\textwidth}
		\centering
		\includegraphics[height=1.5in]{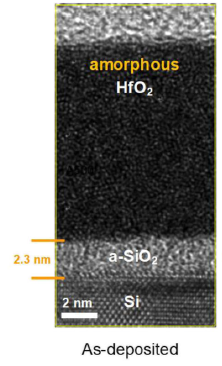}
		\caption{{}}
	\end{subfigure}
	\begin{subfigure}[b]{0.14\textwidth}
		\centering
		\includegraphics[height=1.5in]{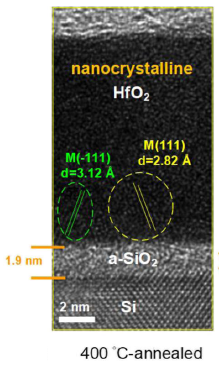}
		\caption{{}}
	\end{subfigure}
	\begin{subfigure}[b]{0.14\textwidth}
		\centering
		\includegraphics[height=1.5in]{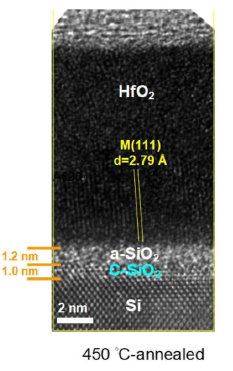}
		\caption{{}}
	\end{subfigure}	
	\begin{subfigure}[b]{0.14\textwidth}
		\centering
		\includegraphics[height=1.5in]{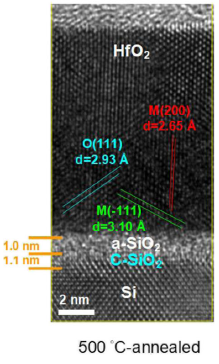}
		\caption{{}}
	\end{subfigure}
	\begin{subfigure}[b]{0.14\textwidth}
		\centering
		\includegraphics[height=1.5in]{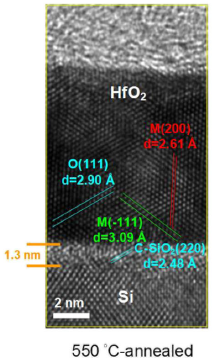}
		\caption{{}}
	\end{subfigure}
	\begin{subfigure}[b]{0.14\textwidth}
		\centering
		\includegraphics[height=1.5in]{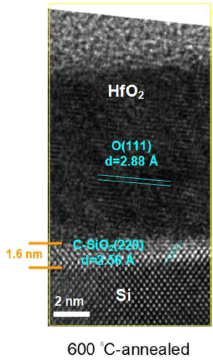}
		\caption{{}}
	\end{subfigure}
	
	\caption{ $HfO_2$ films directly deposited on a $(110)$ substrate using $PE-ALD$ subjected to an oxygen plasma at $2500\;W$ annealed at \textbf{\textit{(a)}} as-deposited\cite{zhang} \textbf{\textit{(b)}} $400^oC$\cite{zhang} \textbf{\textit{(c)}} $450^oC$\cite{zhang} \textbf{\textit{(d)}} $500^oC$\cite{zhang} \textbf{\textit{(e)}} $550^oC$\cite{zhang} \textbf{\textit{(f)}} $600^oC$.\cite{zhang}
	}
\end{figure} 

\begin{figure}[t]
	\centering
	\begin{subfigure}[b]{0.45\textwidth}
		\centering
		\includegraphics[height=1.8in]{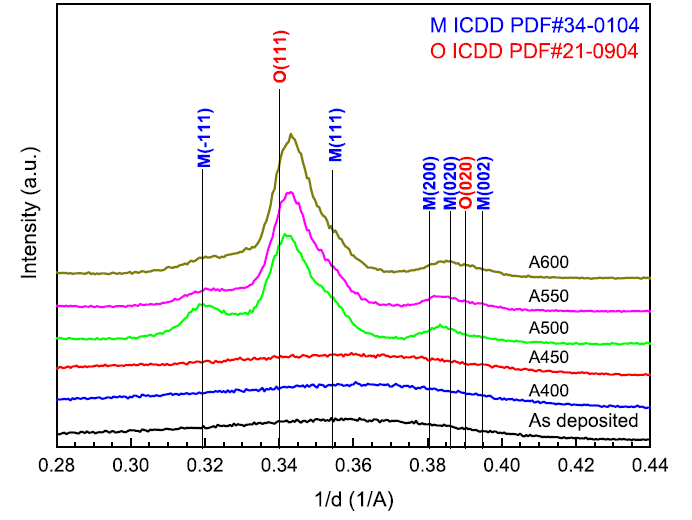}
		\caption{{}}
	\end{subfigure}
	\begin{subfigure}[b]{0.45\textwidth}
		\centering
		\includegraphics[height=1.8in]{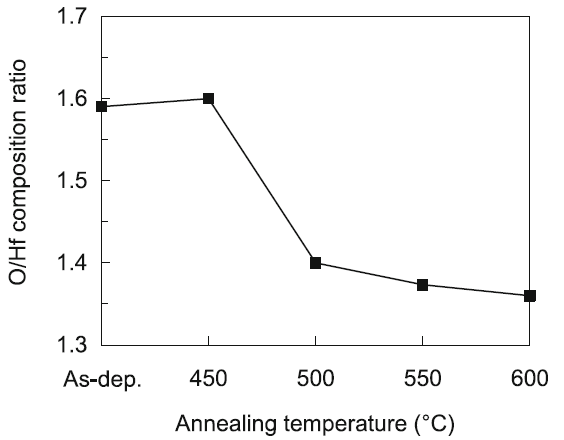}
		\caption{{}}
	\end{subfigure}
	
	\begin{subfigure}[b]{0.20\textwidth}
		\centering
		\includegraphics[height=1in]{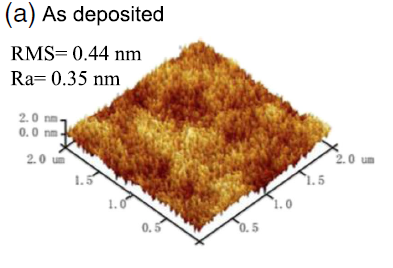}
		\caption{{}}
	\end{subfigure}
	\begin{subfigure}[b]{0.20\textwidth}
		\centering
		\includegraphics[height=1in]{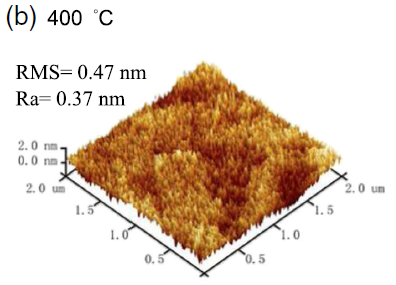}
		\caption{{}}
	\end{subfigure}
	\begin{subfigure}[b]{0.20\textwidth}
		\centering
		\includegraphics[height=1in]{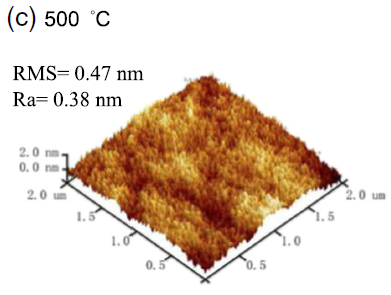}
		\caption{{}}
	\end{subfigure}
	\begin{subfigure}[b]{0.20\textwidth}
		\centering
		\includegraphics[height=1in]{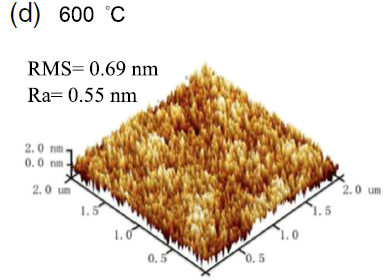}
		\caption{{}}
	\end{subfigure}
	\caption{ \textbf{\textit{(a)}} $GIXRD$ of the samples annealed at different temperatures.\cite{zhang} \textbf{\textit{(b)}} Oxygen composition of the $HfO_2$ layer after being subjected to anneals.\cite{zhang}
	$AFM$ of the sample surface after anneal at a temperature of \textbf{\textit{(c)}} as-deposited\cite{zhang} \textbf{\textit{(d)}} $400^oC$\cite{zhang} \textbf{\textit{(e)}} $500^oC$\cite{zhang} \textbf{\textit{(f)}} $600^oC$.\cite{zhang}
	}
\end{figure} 

\begin{figure}[b]
	\centering
	\begin{subfigure}[b]{0.30\textwidth}
		\centering
		\includegraphics[height=1.1in]{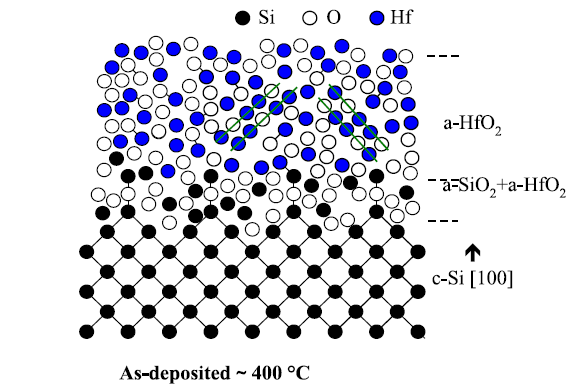}
		\caption{{}}
	\end{subfigure}
	\begin{subfigure}[b]{0.30\textwidth}
		\centering
		\includegraphics[height=1.1in]{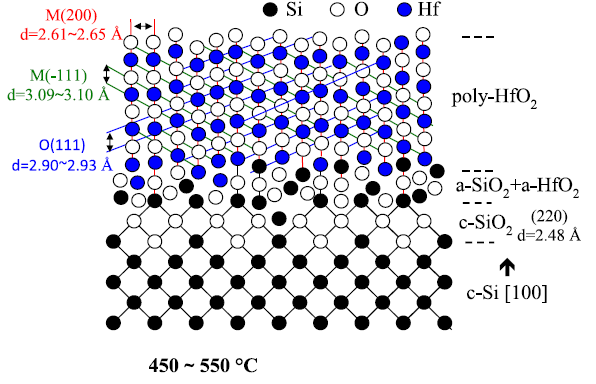}
		\caption{{}}
	\end{subfigure}
	\begin{subfigure}[b]{0.30\textwidth}
		\centering
		\includegraphics[height=1.1in]{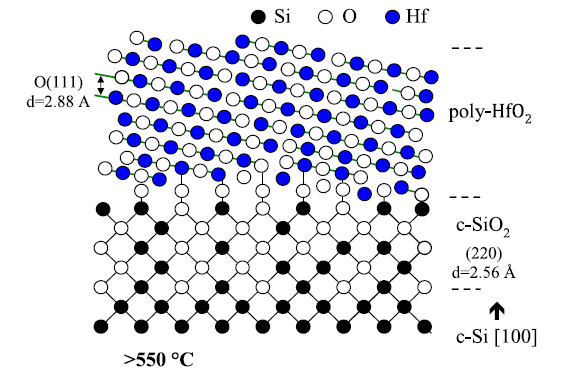}
		\caption{{}}
	\end{subfigure}

	\caption{ Schematic depiction of crystalline arrangements of the dielectric film after different anneal treatments at \textbf{\textit{(a)}} as-deposited to $400^oC$\cite{zhang} \textbf{\textit{(b)}} $450^oC$ to $550^oC$\cite{zhang} \textbf{\textit{(c)}} above $550^oC$.\cite{zhang}
	}
\end{figure}

\begin{figure}[t]
	\centering
\begin{subfigure}[b]{0.45\textwidth}
	\centering
	\includegraphics[height=2in]{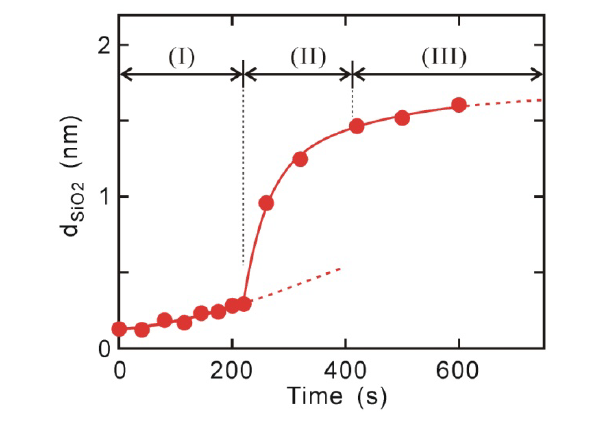}
	\caption{{}}
\end{subfigure}

\begin{subfigure}[b]{1\textwidth}
	\centering
	\includegraphics[height=2in]{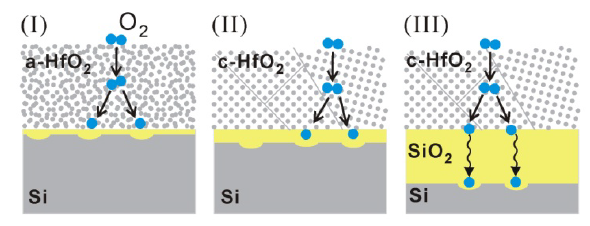}
	\caption{{}}
\end{subfigure}

	\caption{\textbf{\textit{(a)}} Thickness of the interlayer as a function of time in an oxidizing ambient.\cite{miyata} \textbf{\textit{(b)}} Schematic depiction of diffusion mechanism of oxygen atoms at different stages of crystallinity.\cite{miyata}
}

\end{figure}

In order to understand the impetus for the formation of the $SiO_2$ interlayer, one must look at the thermodynamics at the interface.\cite{chang}\cite{gutowski} Thermodynamic stability on silicon is of prime importance for any high-k dielectric, as an unstable interface would promote reaction undesired reactions resulting in a degradation of the device performance.\cite{chang}\cite{gutowski}\cite{giustino} In the case of $HfO_2$ this primarily occurs in the form of an interlayer oxidation.\cite{chang}\cite{gutowski}\cite{giustino} Table 1 shows the various chemical reactions that can occur at the interface and the Gibbs free-energy associated with those reaction pathways at $0K$ and $1000K$.\cite{gutowski}
Gutowski et al, performed $DFT$ simulations of the $ZrO_2/Si$ interface and the $HfO_2$ interface at $0K$ and at $1000K$ and found that $HfO_2$ was much more stable at the interface compared to $ZrO_2$.\cite{gutowski} He also noted that despite the chemical similarities between hafnium and zirconium, due to the higher heat of formation for $HfO_2$ compared to $ZrO_2$ $\sim42\;\frac{KJ}{mol}$ it was more stable at the interface.\cite{gutowski} The lower stability of hafnium silicides compared to zirconium also make it a much more viable option.\cite{gutowski} One can also clearly observe that the formation energy for $SiO_2$ interlayer formation at $0K$ is $227.9\;\frac{KJ}{mol}$, incredibly unfavourable.\cite{gutowski}\cite{shin} That is if $HfO_2$ is deposited on a perfect substrate without any native oxide, no interlayer should form, as observed in Figure 7, where $HfO_2$ films were deposited on silicon substrates freshly treated with $HF$.\cite{gusev} It was also observed that the film quality at the interface in such conditions showed considerable nucleation resulting in non uniform coverage that may cause reliability issues.\cite{gusev}\cite{robertson}\cite{robertson2} Jamison et al, reported a process flow consisting of a $PVD$ step followed by a $RTA$ annealing at $750^oC$, which showed promise in eliminating the interlayer completely with careful control of the anneal times and oxygen flow concentration, as shown in Figure 9.\cite{jamison}$PVD$ and annealing in presence of $O_2$ may prove useful as described in the next section.\cite{jamison}

However, it must be noted that due to incomplete data, no conclusion can be drawn on the mechanism at work.\cite{gutowski} A safe bet for the mechanism of formation at the interlayer is that, most substrates are not free of native oxide.\cite{chang} However thin the oxide may be, it can act as a catalyst for growth due to the negative formation energy for reaction $H7$ forming $HfSiO_4$.\cite{gutowski} 

No analysis of the thermodynamics of the system can be complete without a discussion of the phase diagram of the system shown in Figure 8. \cite{shin}\cite{shin2} Shin et al, calculated the ternary phase diagram and the isopleth of the $Hf-Si-O$ system from the $Hf-Si$, $Si-O$ and  $Hf-O$ binary phase diagrams at $500K$ and $1000K$, as shown.\cite{shin2} From this we can see that the of formation of $HfSiO_4$, with $HfO_2$ and $SiO_2$ as reference states, becomes favourable above a temperature of $1700K$.\cite{shin2} It can also be seen that at $1000K$ $HfO_2$ is stable on the silicon substrate, also observed experimentally.\cite{shin}\cite{gusev} The silicides are known to decompose in a temperature range of $382K-670K$ with the most widely accepted value being $543.5K$, way below most processing temperatures.\cite{chang}\cite{shin2} However, when one observes the $HfO_2-SiO_2$ phase diagram in the same temperature range, one can observe that $HfSiO_4$ is the stable phase, giving credence to our prior proposition for the mechanism of interlayer formation that for a silicate layer to form, the presence of a native oxide layer on the substrate surface is imperative.\cite{gutowski}\cite{shin2}\cite{he}

However, it has been noted that thermodynamic considerations alone cannot explain experimental observations fully.\cite{chang}\cite{zhang} This is because another major factor determining the reactions at the interface is the availability of the reactants, primarily oxygen.\cite{shin}\cite{shin2} Studying the system at different partial pressures therefore is imperative to optimizing any such process. Shin et al, also studies the effects of the local oxygen partial pressure $(p_o)$ on the stability of the phases, when the molar ratio of hafnium to silicon at the interface is set to $1$.\cite{shin}\cite{shin2}
It can clearly be seen that under oxygen deficient ambient the growth of the silicide is promoted at the interface.\cite{kim}\cite{stremmer2} As $p_o$ increases however, the oxide phase becomes stable and is in equilibrium with the silicate, providing us with a narrow but stable window where the interlayer formation can be minimized.\cite{shin}\cite{shin2} Stremmer et al in her extensive review of the subject, investigated in literature the formation of an inter-facial layer in the $HfO_2/Si$ system under different conditions and found that upon increase of $p_o$ from $10^{-4}\;torr$ to $10^{-7}\;torr$ show a significant increase in interlayer thickness.\cite{kim}\cite{stremmer2} They also propose a mechanism involving the diffusion of $SiO$ species at the interface and is discussed in detail in the following sections.\cite{stremmer2}
In the next section we look at the growth of the interlayer and the effect that temperature and other factors have on its thickness and crystallinity.

\subsection*{\normalfont Growth Kinetics}

For any scalable process to be implemented at an industrial scale, the effects of process parameters must be extensively studied and optimized.\cite{chang} Zhang et al, studied the structural evolution of an as-deposited film of $HfO_2$ films using $PE-ALD$ on a p-type silicon substrate under a $2500\;W$ $O_2$ plasma in a $N_2$ ambient.\cite{zhang} It was clearly observed that an interlayer of $SiO_2$ was formed at the interface between the as-deposited $a-HfO_2$ and silicon, which can be attributed to the previous discussion on thermodynamics.\cite{gutowski}\cite{zhang} The films were annealed at a temperature of $400^oC$, and one can clearly observe from the $GIXRD$ data that nano-crystallites of $m-HfO_2$ begin to form which in turn causes an increase in the surface roughness of the film measured using $AFM$.\cite{zhang} $HRTEM$ of the interface also shows that there is no change in the morphology of the interlayer which remains amorphous as shown in Figure 10 and 11.\cite{zhang}
Increasing the anneal temperature to $450^oC$, one can start observing the transition from a fully amorphous interlayer to a partially crystalline one.\cite{zhang} A clear crystallization front can be seen propagating from the substrate upwards whose thickness increases with an increase in anneal temperatures.\cite{zhang} At a $550^oC$ anneal one can observe a complete crystallization of the inter-facial layer, which rearranges itself into a cubic phase with a lattice spacing of $2.48$\AA{ } which increases to $2.56$\AA{ } at $600^oC$.\cite{zhang} This is thought to be influenced by the migration of $O$ atoms from the $HfO_2$ layer to the inter-facial oxide at these temperatures, and is corroborated by the measurements that indicate an oxygen deficient $HfO_2$ layer and an oxygen rich $SiO_2$ layer.\cite{zhang}\cite{copel}\cite{mueller} Such a mechanism is also in agreement with the observation of an increasingly oxygen deficient dielectric as the anneal temperatures are increased.\cite{zhang} At such temperatures the $HfO_2$ exhibits a polycrystalline orthorhombic phase with decreasing lattice spacing with increasing anneal temperatures.\cite{zhang} The maximum dielectric constant of $\sim17.2$ for the film was obtained at $500^oC$, with a decrease at higher anneal temperatures.\cite{zhang} This is presumed to be due to changes in the crystalline arrangement of the inter-facial layer which increases in thickness from $1.3\;nm$ to $1.6\;nm$.\cite{zhang}

Growth under a reducing ambient has also been studied extensively to prevent the formation of the interlayer, with mixed results.\cite{stremmer2} Although an interface that is free of $SiO_2$ has been achieved this usually comes at the cost of a highly oxygen deficient high-k layer, resulting in degradation of film properties.\cite{robertson}\cite{robertson2} Primarily, the electrical properties of the film such as breakdown are degraded and there is a thermodynamic shift towards silicide formation, leading to overall chemical instability.\cite{stremmer2} Thus, it has become a necessary criteria to allow the oxidation of the high-k layer without oxidizing the underlying silicon.\cite{chang}\cite{robertson} In the next section we shall explore the atomistic kinetics at play and look at primarily the mechanism of oxygen diffusion in the lattice and ways to mitigate it at the interface.

\subsection*{\normalfont Atomistic Kinetics \& Diffusion}

To enable the growth of an electrically well behaved $HfO_2$ film, we have established in the previous sections that the growth needs to be performed in an oxidizing environment.\cite{chang}\cite{zhang}\cite{stremmer2} This however comes with a trade-off, growth performed in an oxygen rich environment will inevitably result in the formation of an interlayer.\cite{chang}\cite{stremmer2} Depending on the processing parameters used this interlayer might be a $HfSiO_4$ layer or a $SiO_2$ layer, most commonly the latter.\cite{chang}\cite{zhang}\cite{shin}\cite{he} To prevent the formation of such a layer while maintain the reliability of the dielectric film grown, it is imperative to understand the mechanism of oxygen diffusion in $HfO_2$.\cite{chang}\cite{mckenna}\cite{mueller} The kinetics of such a diffusion will inevitably involve defects in the dielectric film, therefore these are also explored in this section.\cite{mckenna}\cite{mueller}

More often that not, $HfO_2$ films grown on silicon substrates are polycrystalline, even as-deposited films that are initially amorphous result in polycrystalline films when subjected to annealing.\cite{chang}\cite{jamison}\cite{kim3} It is well known that the vacancies in such $HfO_2$ films have a low diffusion activation energy of $\sim0.7\;eV$ in the bulk.\cite{zafar}\cite{mckenna}\cite{mueller} Such defects however, tend to accumulate at the grain boundaries of the deposited film, resulting in diffusion pathways for incoming oxygen atoms.\cite{chang}\cite{almeida}\cite{zafar} Such diffusion behaviour has been studied extensively in literature and has been corroborated by experimental findings.\cite{chang}\cite{almeida}\cite{zafar} The value for nanoscale $HfO_2$ films grown at $p_o=200\;mbar$ is even lower at $\sim 0.52\;eV$, with a diffusion coefficient estimated around $\sim10^{14}\;cm^{2} s^{-1}$.\cite{mckenna}\cite{zafar} First principle studies of positive oxygen vacancies in $HfO_2$ are in excellent agreement with these results.\cite{caymax}\cite{mckenna} The mechanism therefore is at least partially dependent on the diffusion of oxygen from the oxidizing ambient through positive vacancies that accumulate at the grain boundaries.\cite{zafar}\cite{stremmer2} It has also been proposed that $Si^{4+}$ and $SiO$ intermediate species and their diffusion might also have a role to play at the interface.\cite{stremmer2} Under moderately reducing environments, such as those created by forming gas, it has been found that the activity of these species is suppressed resulting in reduced or non-existent interlayer formation.\cite{zafar}\cite{stremmer2} In any case, it is evident that the primary cause for interlayer formation is not only the direct reaction between the $HfO_2$ and silicon at the interface but diffusion of oxygen from the oxidizing ambient through the defects at the grain boundaries to the interface.\cite{mckenna}\cite{zafar} Now that we can pinpoint the cause of interlayer formation, we are in good shape to come up with measures to mitigate its formation.

\subsection*{\normalfont Interlayer Prevention}

\begin{figure}[t]
	\centering
	\begin{subfigure}[b]{0.45\textwidth}
		\centering
		\includegraphics[height=1.6in]{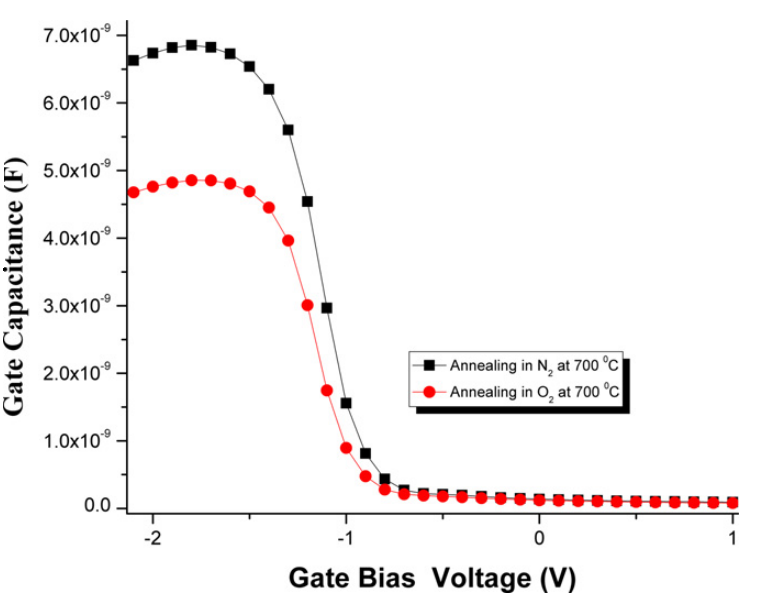}
		\caption{{}}
	\end{subfigure}
	\begin{subfigure}[b]{0.45\textwidth}
		\centering
		\includegraphics[height=1.6in]{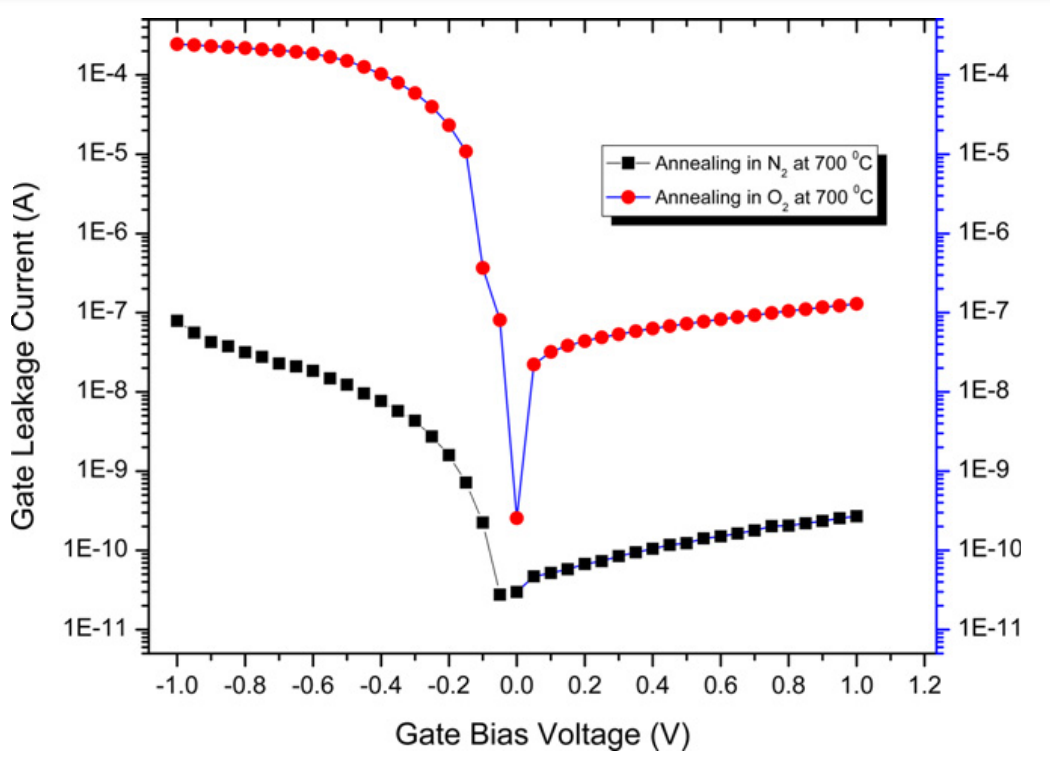}
		\caption{{}}
	\end{subfigure}

	\begin{subfigure}[b]{0.45\textwidth}
		\centering
		\includegraphics[height=1.6in]{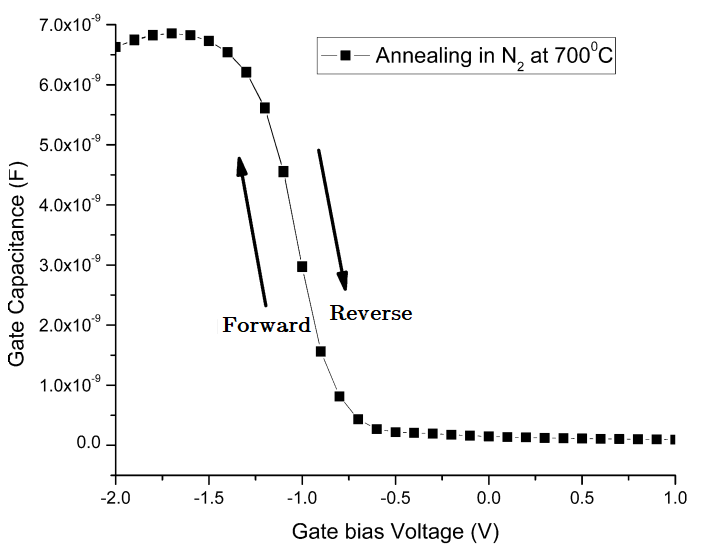}
		\caption{{}}
	\end{subfigure}
	\begin{subfigure}[b]{0.45\textwidth}
		\centering
		\includegraphics[height=1.6in]{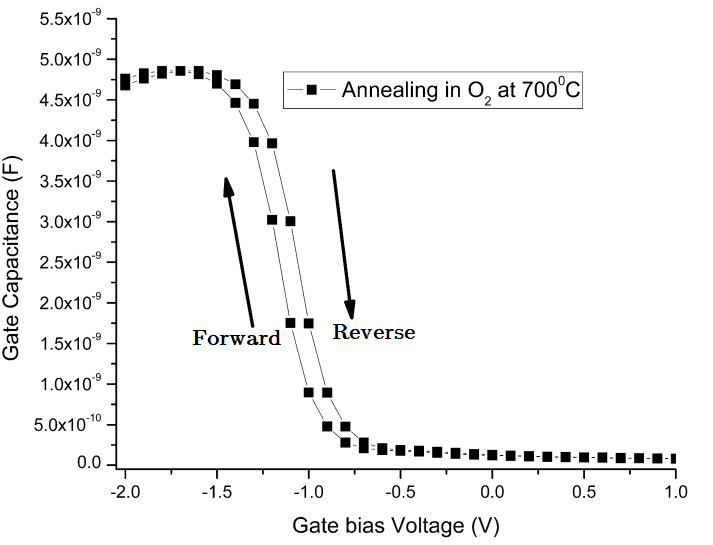}
		\caption{{}}
	\end{subfigure}

	\caption{ $C-V$ traces and leakage currents of $HfO_2$ films grown in an oxygen ambient compared to films grown in a nitrogen ambient.
	}
\end{figure} 

Given the mechanism of the diffusion which forms the oxide interlayer, the primary approach to mitigate its formation is crystal clear. If the positive vacancies at the grain boundaries that enable diffusion are occupied by another less diffusive species during growth, oxygen can react with the $HfO_2$ fully oxidizing it but will not be able migrate to the interface.\cite{robertson}\cite{robertson2} Narayanan et al, show this by utilizing nitrogen as the growth ambient which has a tendency to accumulate at the grain boundaries, preventing the reacting oxygen from reaching the interface in $Y_2O_3$, with similar studies also done on $HfO_2$ by other groups.\cite{narayanan2}\cite{narayanan} This has an added advantage of forming dielectric films with a lower defect density, better surface morphology and interface quality, thus, making it a popular approach.\cite{narayanan} Nitridation has an added benefit of forming oxy-nitrides instead of pure oxides which are well known to have a higher dielectric constant, thus, better $C-V$ performance as seen in Figure 14.\cite{singh}\cite{narayanan}\cite{wong}

It is also possible to completely eliminate the interlayer by selectively desorbing the $SiO_2$ on the interface during growth by annealing at above $1000K$, this is also known to happen in buried layers.\cite{robertson} It has also been demonstrated that addition of metallic hafnium at the interface also causes decomposition of the interlayer due to $Hf$ displacing the $Si$ in the $SiO_2$ lattice.\cite{robertson} It should however be noted that, as long as the thickness of the inter-facial oxide can be controlled, it is usually not completely destroyed due to a couple of reasons. Keeping the interlayer also has advantages that make further growth and processing more convenient.\cite{chang}\cite{robertson}

$HfO_2$ films grown using $ALD$ are usually grown on substrates with a thin layer of native oxide, this is because the presence of a native oxide provides better adhesion of the precursors to the surface.\cite{chang}\cite{robertson} This acts as a nucleating layer for the growth.\cite{chang}\cite{robertson} It also improves the quality of the interface between the high-k dielectric and the silicon, resulting in improvements in its electrical behaviour.\cite{chang}\cite{robertson} The $SiO_2/Si$ interface is well understood and if made of high quality and low defect layer, can help mitigate the lowering of the channel mobility that most high-k dielectrics cause.\cite{robertson} Recent research also suggests that the electrical behaviour of the interlayer deviates from bulk $SiO_2$, showing a lower reduction in $EOT$ than what was expected.\cite{giustino}

In the next section we shall take a closer look into the effects that the interlayer has on device performance and critically discuss its advantages and disadvantages.

\section*{\normalfont Effects on CMOS Device Parameters \& Performance}

\begin{figure}[t]
	\centering
	\begin{subfigure}[b]{0.45\textwidth}
		\centering
		\includegraphics[height=1.2in]{fig7_1.png}
		\caption{{}}
	\end{subfigure}
	\begin{subfigure}[b]{0.45\textwidth}
		\centering
		\includegraphics[height=1.2in]{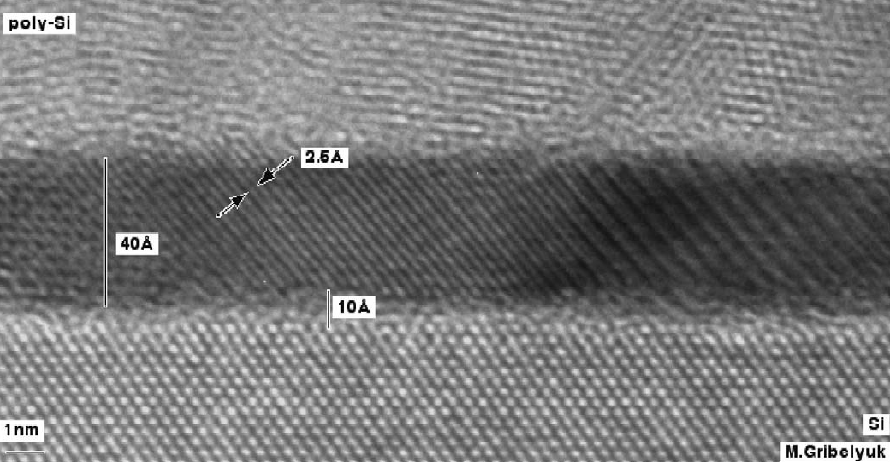}
		\caption{{}}
	\end{subfigure}
	
	\begin{subfigure}[b]{0.45\textwidth}
		\centering
		\includegraphics[height=1.6in]{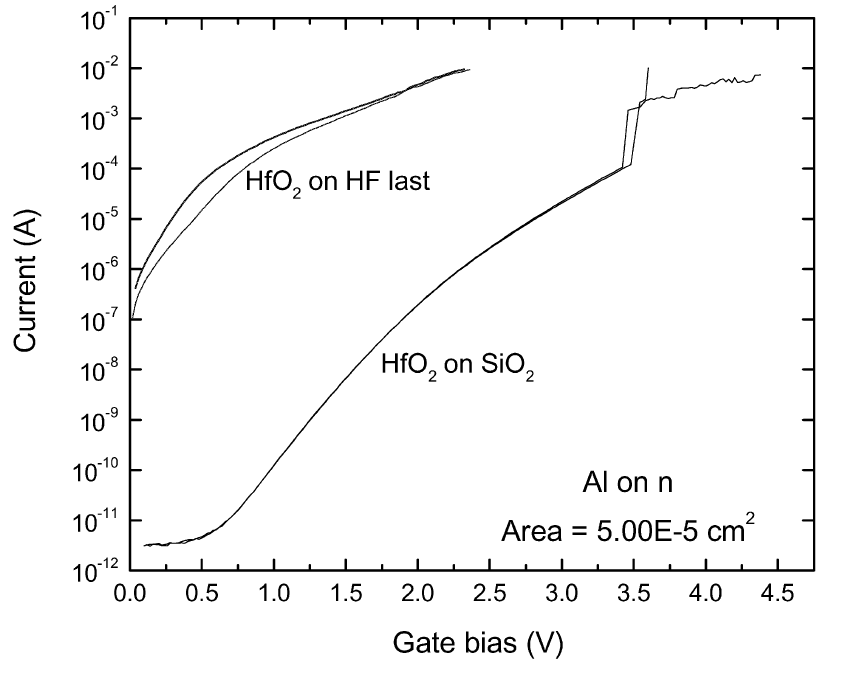}
		\caption{{}}
	\end{subfigure}
	\begin{subfigure}[b]{0.45\textwidth}
		\centering
		\includegraphics[height=1.6in]{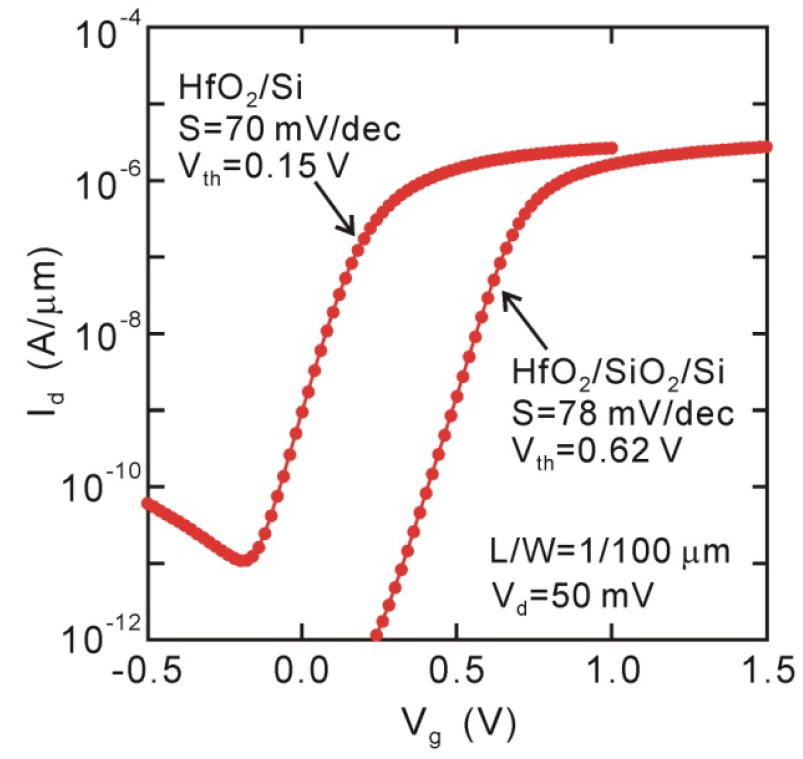}
		\caption{{}}
	\end{subfigure}

	\caption{ \textit{\textbf{(a)}} $HRTEM$ image of an abrupt $HfO_2/Si$ interface.\cite{gusev}
		\textit{\textbf{(b)}} $HRTEM$ image of a $HfO_2/SiO_2-40$\AA$/Si$ interface.\cite{gusev}
		\textit{\textbf{(c)}} $I_{d}-V_{g}$ characteristics of a $HfO_2/Si$ device compared to a $HfO_2/SiO_2-40$\AA$/Si$ device.\cite{gusev} \textit{\textbf{(d)}} A $HfO_2/SiO_2/Si$ device compared to an aligned dipole $HfO_2/SiO_2/Si$ device subjected to a post deposition anneal.\cite{miyata}
	}
\end{figure} 

The primary motivation for a deeper understanding of the mechanism of interlayer formation in the $HfO_2-Si$ gate stack was to understand its impact on the performance of next generation $CMOS$ devices. A keen understanding of this phenomenon was crucial in weighing trade-off between performance enhancement and process limitations, that is if it was desirable to eliminate the interlayer at the cost of a poorer interface. In this section we look at the impact that the presence of an interlayer has on the $CMOS$ performance factors of the device and the advantages and disadvantages that it entails.

\subsection*{\normalfont $I_{d}-V_{g}$ Characteristics}

The primary quantity controlling the overall behaviour of any $MOSFET$ is the capacitance at the gate $C_{g}$, which ultimately is dependent on the effective thickness of the oxide.\cite{robertson} In a gate stack consisting of only $HfO_2$ and silicon without an interlayer, one would expect the capacitance to be contributed only by the high-k oxide.\cite{robertson} In reality rarely does this occur in real devices, where the interlayer acts as a parasitic capacitance $(C_i)$ in series with the oxide capacitance $(C_{ox})$.\cite{robertson} This lowers the effective dielectric constant of the stack, therefore lowering gate control over the channel, observed experimentally.\cite{robertson}\cite{chiu} This has a direct consequence in altering the transfer characteristics of the device, driven directly by the efficiency of channel formation and strong inversion.\cite{robertson}\cite{chiu} Figure 15 compares the transfer characteristics of two such devices one with an interlayer and the other without.\cite{gusev}\cite{miyata} It can clearly be seen that the maximum current in the device with the interlayer is reduced significantly in the linear region.\cite{gusev} The saturation region in the device with an ultra-thin $40$\AA{ } interlayer is also significantly degraded but is pointed to be due to the increased surface trap density of the interlayer oxide formed.\cite{gusev} The devices measured in Figure 15(d) show that given a good enough interface quality the $HfO_2/SiO_2/Si$ stack has the ability to perform as well as a direct contact device.\cite{miyata} 

\subsection*{\normalfont Mobility}

\begin{figure}[t]
	\centering
	\begin{subfigure}[b]{0.3\textwidth}
		\centering
		\includegraphics[height=1.4in]{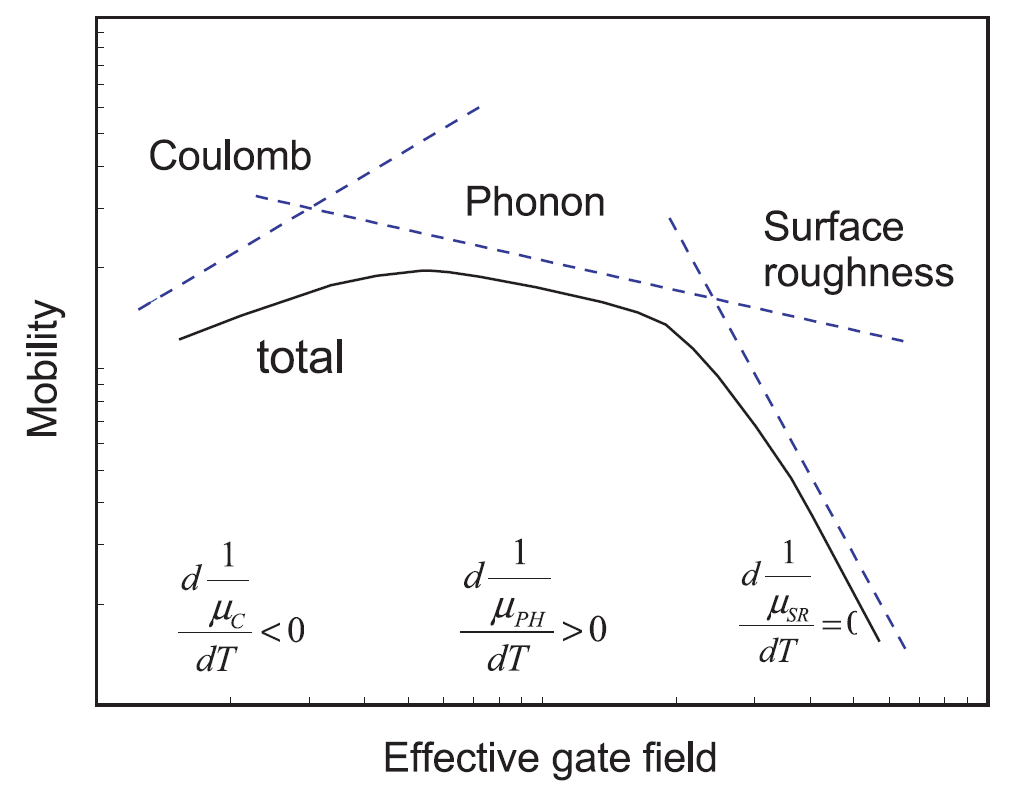}
		\caption{{}}
	\end{subfigure}
	\begin{subfigure}[b]{0.3\textwidth}
		\centering
		\includegraphics[height=1.4in]{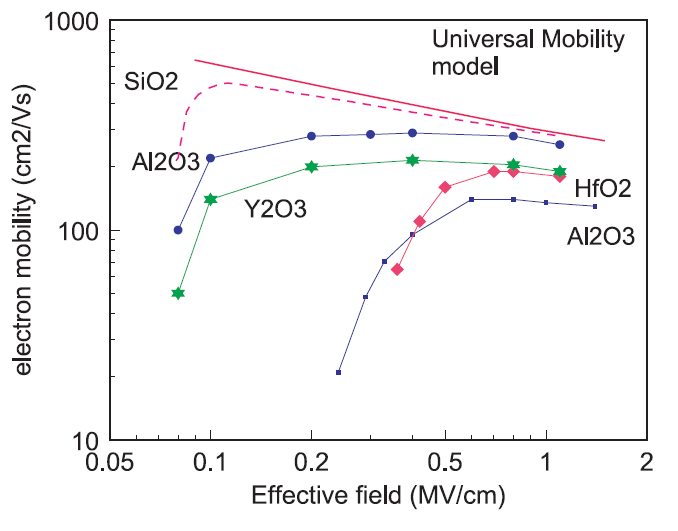}
		\caption{{}}
	\end{subfigure}
	\begin{subfigure}[b]{0.3\textwidth}
		\centering
		\includegraphics[height=1.4in]{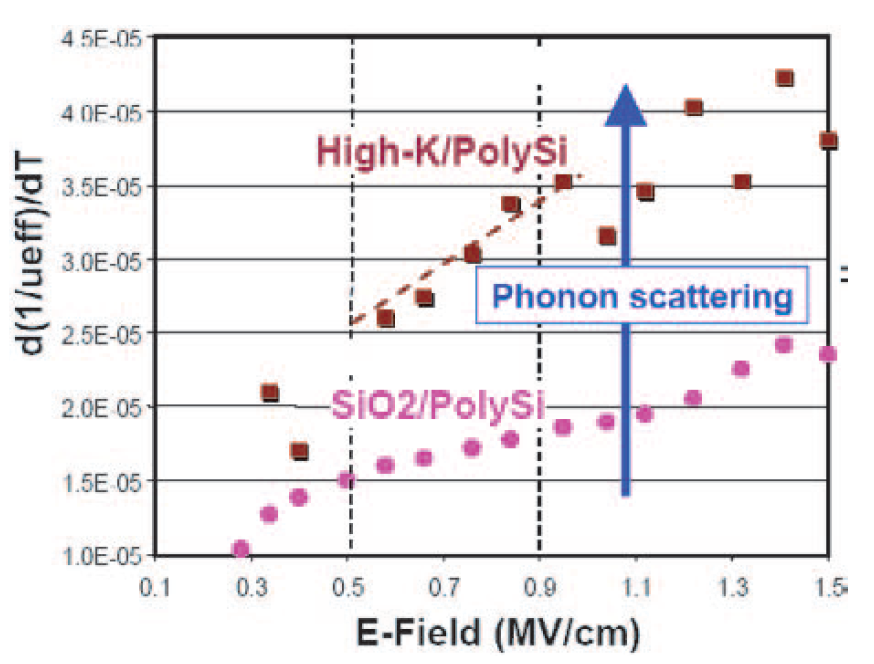}
		\caption{{}}
	\end{subfigure}

	\begin{subfigure}[b]{0.30\textwidth}
		\centering
		\includegraphics[height=1.4in]{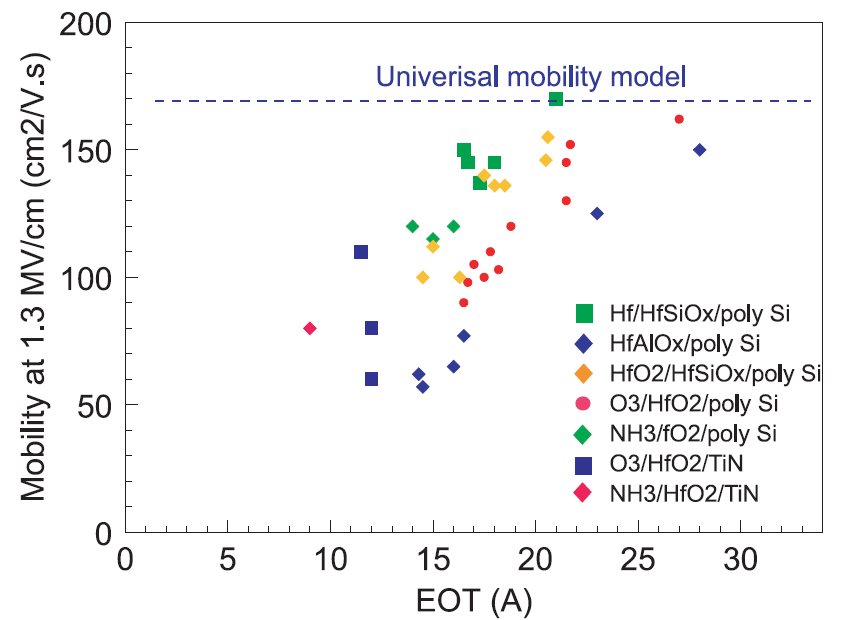}
		\caption{{}}
	\end{subfigure}
	\begin{subfigure}[b]{0.30\textwidth}
		\centering
		\includegraphics[height=1.4in]{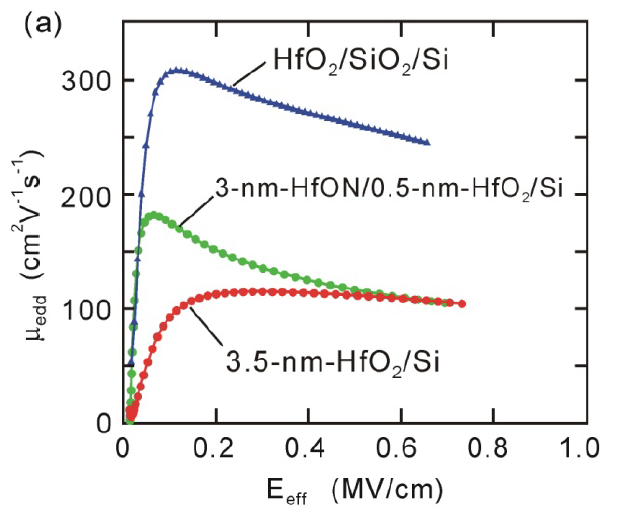}
		\caption{{}}
	\end{subfigure}
	\begin{subfigure}[b]{0.30\textwidth}
		\centering
		\includegraphics[height=1.4in]{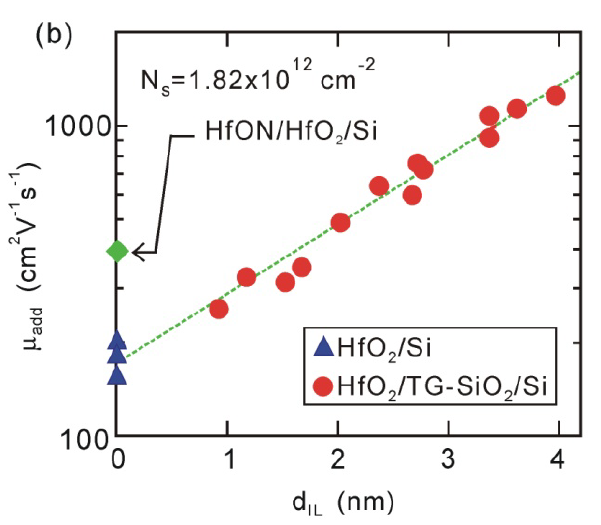}
		\caption{{}}
	\end{subfigure}

	\caption{ \textbf{\textit{(a)}} Schematic depiction of mobility reduction mechanisms at different oxide fields.\cite{robertson} \textbf{\textit{(b)}} Mobility reduction according to  the universal mobility model.\cite{robertson} \textbf{\textit{(c)}} Change in the mobility with temperature as a function of field.\cite{robertson} \textbf{\textit{(d)}} Mobility reduction as a function of $EOT$.\cite{robertson}\textbf{\textit{(e)}} Electron mobilities in a $3-nm\;HfO_2/Si$ stack,$3.5-nm\;HfO_2/Si$ stack and a $HfO_2/SiO_2/Si$ stack.\cite{miyata} \textbf{\textit{(f)}} Added electron mobility as a function of interlayer thickness.\cite{miyata}
	}
\end{figure}

The primary reason for avoiding a direct interface between the $HfO_2$ and the silicon is not the $EOT$ loss, but also the severe degradation in the mobility of carriers at the channel in such devices.\cite{robertson} One can clearly see the degradation in the mobility in a $HfO_2/Si$ stack compared to a $HfO_2/SiO_2/Si$ stack, shown in Figure 16.\cite{robertson}\cite{miyata} $CMOS$ devices with a $SiO_2$ gate dielectric have a channel mobility close to the universal limit derived from Poisson's equation.\cite{robertson} Any degradation in the channel mobility is the result of a scattering mechanism, and is accounted for in the effective mobility of the carriers in the channel given by Matthiessen’s rule.\cite{robertson} As a rule of thumb, at low fields, the scattering is limited by coulombic interactions between carriers and trapped charges.\cite{robertson} While at moderate and high fields they are primarily limited by phonon interactions and surface roughness respectively.\cite{robertson}\cite{miyata} The exact mechanism of this reduction in mobility is hard to pinpoint though several promising contenders exist.\cite{robertson} Most of the literature agrees that high-k oxides in general have a larger density of trapped charges in them, which in turn results in the carriers experiencing excessive scattering.\cite{robertson} Another mechanism proposed for the lower mobility in high-k dielectrics is carrier scattering by low frequency polar lattice vibration modes which are also the primary cause of a high dielectric constant. The dilemma arises when one realizes that this is a direct trade-off between $EOT$ and mobility, making it an inevitable compromise.\cite{robertson} The dependence of the mobility reduction on both the temperature and the thickness of the entire gate stack indicate that both mechanism's, coulombic and phonon scattering are operative.\cite{robertson} Devices grown with an interlayer only show a marginally lower mobility than those with $SiO_2$ gate oxides, indicating the remote scattering mechanism at play.\cite{robertson} Chau et al, suggests that this may be overcome by the screening of interface dipoles by metal electrodes, providing a concrete impetus for metal/high-k stacks.\cite{robertson} The use of a nitrogen ambient to partially nitride our oxide to create a hybrid $HfON/HfO_2/Si$ stack has shown significant improvement in interface quality, leading to a significant increase in mobility.\cite{robertson}\cite{narayanan}

\subsection*{\normalfont Threshold Voltage}

\begin{figure}[t]
	\centering
	\begin{subfigure}[b]{0.45\textwidth}
		\centering
		\includegraphics[height=1.6in]{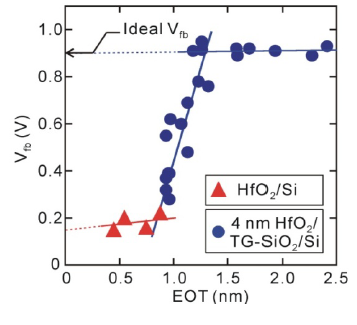}
		\caption{{}}
	\end{subfigure}
	\begin{subfigure}[b]{0.45\textwidth}
		\centering
		\includegraphics[height=1.6in]{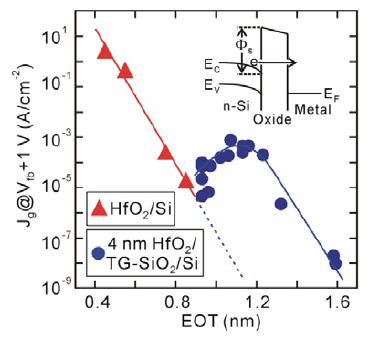}
		\caption{{}}
	\end{subfigure}

	\begin{subfigure}[b]{0.45\textwidth}
		\centering
		\includegraphics[height=1.6in]{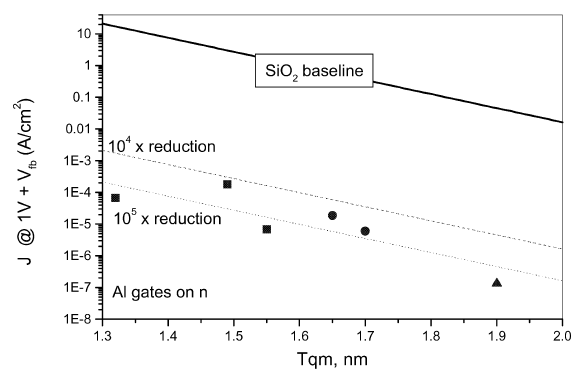}
		\caption{{}}
	\end{subfigure}
	\begin{subfigure}[b]{0.45\textwidth}
		\centering
		\includegraphics[height=1.6in]{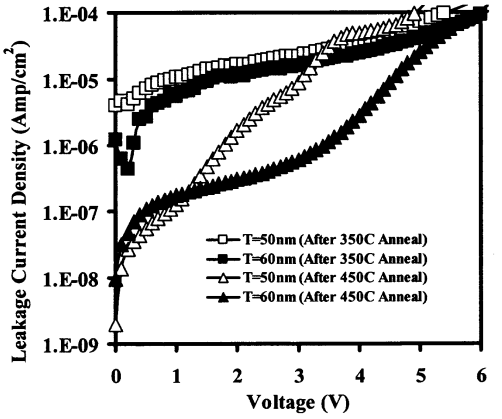}
		\caption{{}}
	\end{subfigure}

	\caption{ \textbf{\textit{(a)}} Flat band voltage as a function of $EOT$ in a direct contact device and an interlayer device.\cite{miyata} \textbf{\textit{(b)}} Leakage current as a function of $EOT$ in a direct contact device and an interlayer device.\cite{miyata} \textbf{\textit{(c)}} Leakage current as a function of $EOT$ in a direct contact device.\cite{gusev} \textbf{\textit{(d)}} Leakage current as a function of thickness and anneal treatment.\cite{garg}
	}
\end{figure}

Another very important parameter that is affected by the formation of an interlayer is the gate potential at which strong inversion is initiated.\cite{robertson}\cite{garg} This is commonly termed the threshold voltage of the device and can be characterized both from the transfer characteristics and the $C-V$ trace.\cite{miyata}  It is immediately noticeable that the threshold of the device without the interlayer is significantly shifted to the left towards $0V$ and sometimes even negative, indicating ease of strong inversion even at low gate biases.\cite{robertson}\cite{miyata} This is usually not desired as it indicates a large flat-band voltage $\sim1\;V$ between the electrode and the oxide as seen in Figure 17.\cite{robertson}\cite{miyata} This is another reason polysilicon is not the preferred gate electrode in such devices due to the need of a large metal work-function to induce inversion.\cite{robertson}\cite{miyata} In silicon, the gate electrode should ideally be able to swing the fermi level by $1.1\;eV$ to be of any use, only possible with metals in high-k devices, and are different depending on whether it is a $PMOS$ or $NMOS$ devices.\cite{robertson}\cite{miyata}\cite{onishi} The presence of an interlayer alleviates this problem a little by lowering the flat-band voltage, presumably by preventing dipole interactions at the interface electrode high-k interface.\cite{robertson}\cite{miyata}\cite{onishi}

Threshold voltage shifts may also be partially attributed to a poor interface oxide resulting in trapped charges at the interface, resulting in the distortion of the $C-V$ characteristics.\cite{robertson}\cite{singh}\cite{garg} Charge trapping at the high-k layer also results in the pinning of the fermi level at the interface and instability of the flat band voltage, and therefore the threshold voltage.\cite{chang}\cite{robertson}\cite{miyata} This effect has been demonstrated clearly by charge pumping experiments, where gate hysteresis in the $C-V$ indicates clear presence of trapped charges as seen in Figure 14.\cite{singh} It can also be clearly observed that oxides annealed or grown in a nitrogen ambient do not exhibit this behaviour.\cite{singh} This provides us with an inkling into the origin of these trapped charges, primarily caused due to oxygen defects, $Hf^{3+}$ ions and oxygen interstitials.\cite{robertson} These can be reduced by repeated annealing of the dielectric in a nitrogen atmosphere or a forming gas ambient.\cite{robertson} This has the added benefit of compacting dielectric layers grown by $ALD$, the primary method used in industry for deposition.\cite{robertson}\cite{chang}

\subsection*{\normalfont Leakage Current}

\begin{figure}[t]
	\centering
	\begin{subfigure}[b]{0.3\textwidth}
		\centering
		\includegraphics[height=1.5in]{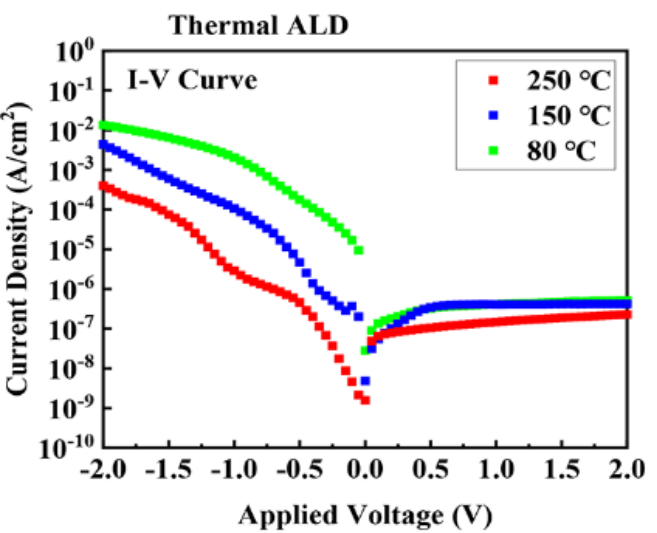}
		\caption{{}}
	\end{subfigure}
	\begin{subfigure}[b]{0.3\textwidth}
		\centering
		\includegraphics[height=1.5in]{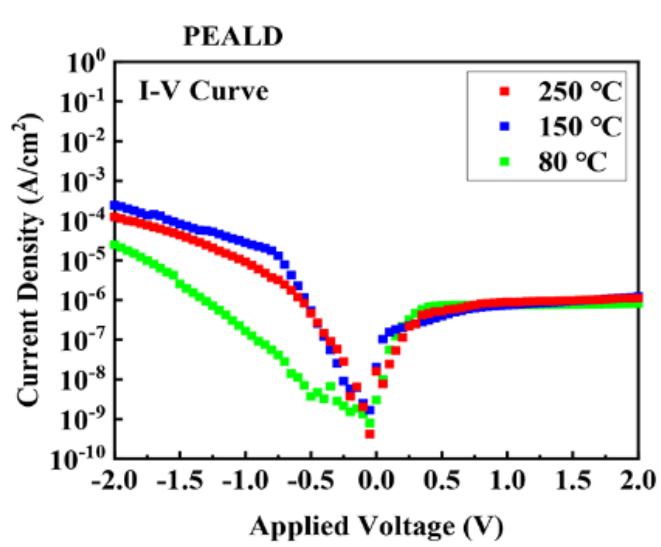}
		\caption{{}}
	\end{subfigure}
	\begin{subfigure}[b]{0.3\textwidth}
		\centering
		\includegraphics[height=1.3in]{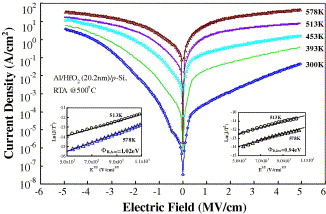}
		\caption{{}}
	\end{subfigure}

	\begin{subfigure}[b]{1\textwidth}
		\centering
		\includegraphics[height=1.3in]{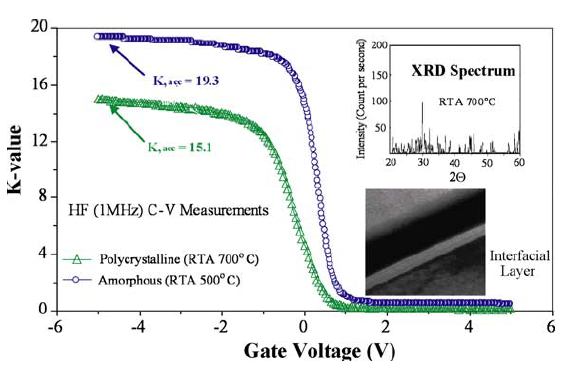}
		\caption{{}}
	\end{subfigure}

	\begin{subfigure}[b]{0.3\textwidth}
		\centering
		\includegraphics[height=1.3in]{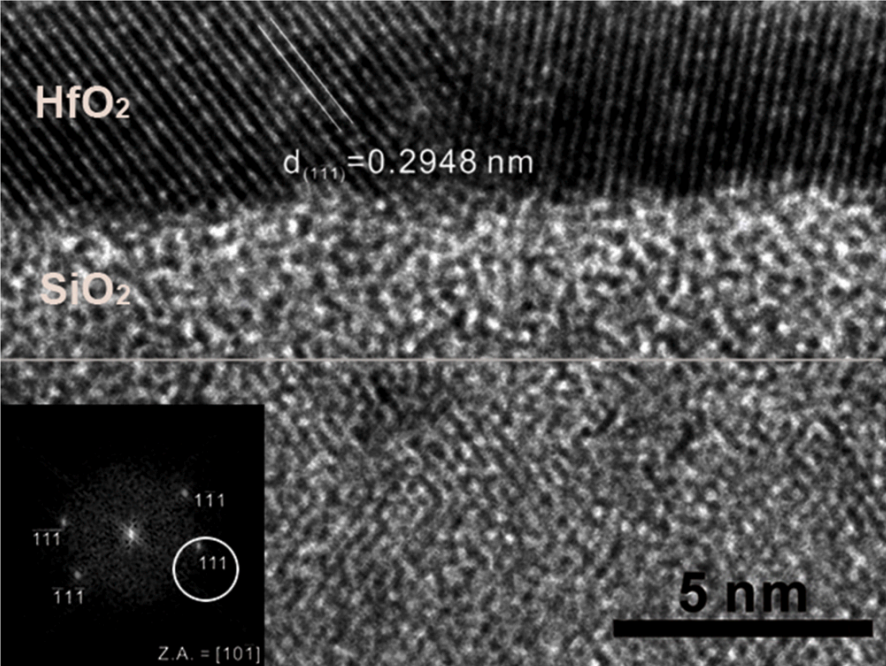}
		\caption{{}}
	\end{subfigure}
	\begin{subfigure}[b]{0.3\textwidth}
		\centering
		\includegraphics[height=1.3in]{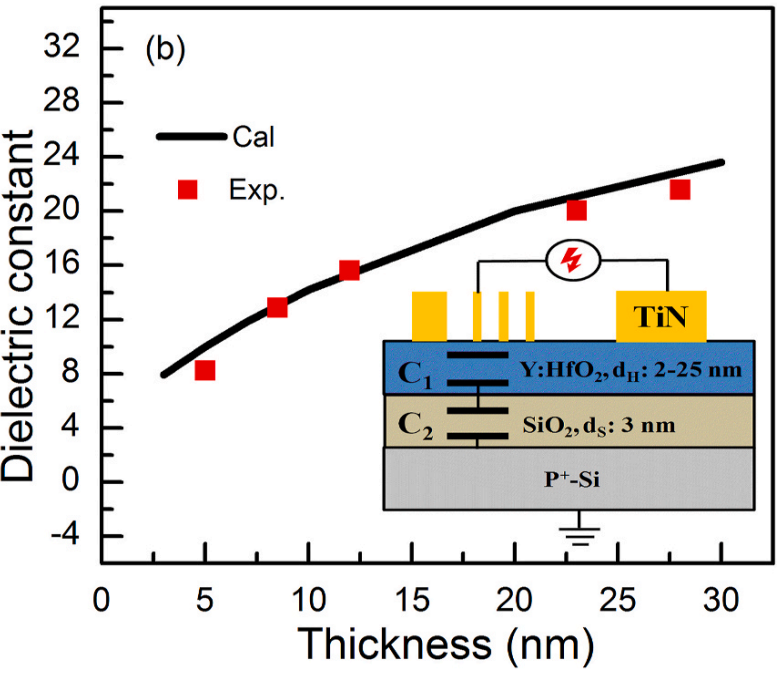}
		\caption{{}}
	\end{subfigure}
	\begin{subfigure}[b]{0.3\textwidth}
		\centering
		\includegraphics[height=1.3in]{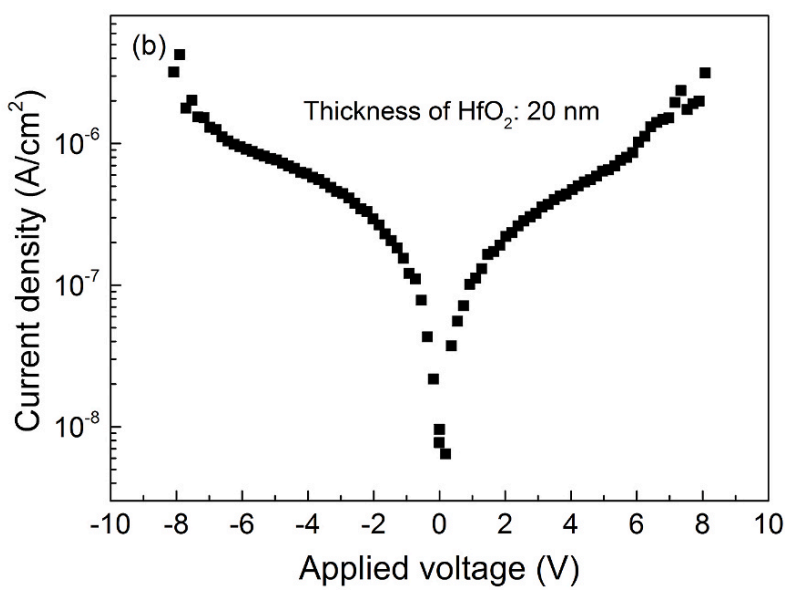}
		\caption{{}}
	\end{subfigure}
	
	\caption{ \textbf{\textit{(a)}} Leakage characteristics of films deposited by Thermal $ALD$.\cite{kim3}
		\textbf{\textit{(b)}} Leakage characteristics of films deposited by $PE-ALD$.\cite{kim3}
			\textbf{\textit{(c)}} Leakage characteristics of films deposited by magnetron-sputtering as a function of temperature.\cite{chiu}
					\textbf{\textit{(d)}} $C-V$ trace of an amorphous $HfO_2$ film compared to the same film, which is now polycrystalline film after annealing at $700^oC$\cite{chiu}
		\textbf{\textit{(c)}} $TEM$ image of the interface in a $TiN/HfO_2/SiO_2-2.88\;nm\;/Si$ $MOSCAP$.\cite{chiu}\cite{sun}
		\textbf{\textit{(d)}} Dielectric constant as a function of thickness.\cite{sun}
		\textbf{\textit{(e)}} Gate leakage characteristics of  a $20\;nm\; HfO_2$ film.\cite{sun} 	}
\end{figure}

In the case of a $HfO_2/SiO_2/Si$ gate stack with an interlayer it is expected that the gate leakage current would be reduced due to an increase in the thickness of the oxide.\cite{chang}\cite{robertson}\cite{caymax} However, the physical thickness of the oxide is not the only factor on which the leakage current is dependent, with the homogeneity of the crystalline phase; as well as the trap density playing an important role.\cite{gusev}\cite{gougam}\cite{chiu}\cite{giustino} Thicker films as expected show a lower leakage current at low biases, although this difference becomes negligible as the bias is increased.\cite{chiu}\cite{garg} The gate current is also largely dependent on the defect density in the material and an appropriate anneal cycle can reduce gate leakage by orders of magnitude, as shown in Figure 17.\cite{chiu}\cite{kim3} This is presumed to be due to defect annihilation and passivation of the traps upon annealing.\cite{kim3}\cite{mckenna} Annealing in a nitrogen ambient has also shown much promise in decreasing the leakage current as shown in Figure 14.\cite{singh}\cite{narayanan}\cite{narayanan2} This is most likely due to partial nitridation of the $HfO_2$ as well as the interlayer.\cite{robertson}\cite{singh}\cite{narayanan2}\cite{narayanan} Increasing the oxygen flow duration during the growth of the high-k layer has also shown promise, most likely due to the annihilation of the positive oxygen vacancies at the grain boundaries, that act as electron conduction pathways leading to leakage in the first place.\cite{banyay}\cite{zhang} As deposited films also show much better leakage characteristics compared to annealed films, presumably due to their amorphous nature as observed in Figure 18, where films deposited using $PEALD$ showed significantly better switching characteristics than those deposited using Thermal $ALD$.\cite{zhang}\cite{kim3} The degradation in the leakage characteristics of the $PEALD$ films is hypothesized to be due to the formation of polycrystalline domains in a disordered lattice, leading to defect migration and formation of leakage pathways.\cite{almeida}\cite{mckenna} In contrast the improvement in the polycrystalline films deposited by thermal $ALD$ are presumed to be due to defect annihilation but further evidence is required.\cite{chiu}\cite{kim3}

\section*{\normalfont Conclusion}
In this term paper we looked at and analysed an in depth mechanism of interlayer formation in the $HfO_2/Si$ system at the interface and provide a thermodynamic as well as kinetic impetus for such behaviour. The structural and electrical properties of $HfO_2$ were explored briefly after which thermodynamic argument was presented.\cite{chang} It was argued that the relative stability of the deposited $HfO_2$ and the highly positive Gibbs energy $(\Delta G_0=227.9\;\frac{KJ}{mol})$ for such a system could not be the root cause, hence motivating an atomistic description of the process. A detailed look at the phase diagram also showed us that the interface was relatively stable upto $\sim 1700K$ much larger than any process temperatures.\cite{gutowski}\cite{shin2} This points the root cause of interlayer formation to a different source, namely the partial pressure of the oxygen in the process environment.\cite{shin}\cite{shin2} Oxygen is a necessary component in maintaining the integrity of the film, thus, is a necessary evil.\cite{stremmer2} It was found that oxygen diffusion along the grain boundaries of the polycrystalline film were the primary culprit.\cite{mckenna} This is primarily due to the positive vacancies in the $HfO_2$ aiding oxygen diffusion along the grain boundaries.\cite{chang}\cite{mckenna} Additionally, the diffusion of the fast $SiO$ species at the interface is also thought to be a contributing factor.\cite{stremmer2} This is clearly observed in oxidizing environments where the thickness of the interlayer increases on increased annealing in oxygen. We also explored some of the steps to mitigate this by the growth in a nitrogen ambient preventing diffusion of oxygen and resulting in thinner or non-existent interlayers.\cite{stremmer2} Careful annealing was also explored, though briefly.\cite{jamison}

The industry seems to have agreed that the interlayer is a necessary evil, on one hand, it clearly increases the $EOT$ of the device, on the other hand, it proves a much needed increase in mobility not found in direct contact devices.\cite{robertson}\cite{lime} Such devices also show higher flat band voltage, thus, a rightward shift in their threshold voltage, usually undesired for proper scaling in $CMOS$ technology.\cite{robertson}\cite{miyata} The gate leakage characteristics could also be controlled by controlling the thickness of the interlayer.\cite{robertson}\cite{wang} This is achievable through good process control, leading to a negligible reduction in dielectric constant even for very thin high-k films.\cite{chang}
In conclusion, the industry has figured out a magic formula to perfect the interface between high-k dielectrics and silicon, and while it is most likely not $HfO_2$ it is important to remember that $HfO_2$ was the material that pushed moore's law into the $21'st$ century making further innovation possible.\cite{james} Current literature seems to be focused on other more exotic hafnium materials like aluminates and lanthanates for future scaling, but to understand the mechanism of interlayer formation is still a fruitful endeavour.\cite{he2}


\bibliography{myref}
\bibliographystyle{unsrt}

\end{document}